\documentclass[twocolumn]{aastex631}

\defcitealias{LustigYaeger2023}{Lustig-Yaeger \& Fu et al. 2023}
\defcitealias{Moran2023}{Moran \& Stevenson et al. 2023}
\defcitealias{May2023}{May \& MacDonald et al. 2023}

\begin{document}

\title{JWST COMPASS: A NIRSpec/G395H Transmission Spectrum of the Sub-Neptune TOI-836c} 

\author[0000-0003-0354-0187]{Nicole L. Wallack} 
\affiliation{Earth and Planets Laboratory, Carnegie Institution for Science, Washington, DC 20015, USA}

\author[0000-0003-1240-6844]{Natasha E. Batalha}
\affiliation{NASA Ames Research Center, Moffett Field, CA 94035, USA}

\author[0000-0001-8703-7751]{Lili Alderson} 
\affiliation{School of Physics, University of Bristol, HH Wills Physics Laboratory, Tyndall Avenue, Bristol BS8 1TL, UK}

\author[0000-0003-3623-7280]{Nicholas Scarsdale}
\affiliation{Department of Astronomy and Astrophysics, University of California, Santa Cruz, CA 95064, USA}

\author[0000-0002-4489-3168]{Jea I. Adams Redai} 
\affiliation{Center for Astrophysics ${\rm \mid}$ Harvard {\rm \&} Smithsonian, 60 Garden St, Cambridge, MA 02138, USA}

\author[0000-0002-8949-5956]{Artyom Aguichine} 
\affiliation{Department of Astronomy and Astrophysics, University of California, Santa Cruz, CA 95064, USA}

\author[0000-0003-4157-832X]{Munazza K. Alam}
\affiliation{Space Telescope Science Institute, 3700 San Martin Drive, Baltimore, MD 21218}

\author[0000-0002-8518-9601]{Peter Gao} 
\affiliation{Earth and Planets Laboratory, Carnegie Institution for Science, Washington, DC 20015, USA}

\author[0000-0003-2862-6278]{Angie Wolfgang}
\affiliation{Eureka Scientific, Inc.
2452 Delmer Street Suite 100
Oakland, CA 94602-3017}

\author[0000-0002-7030-9519]{Natalie M. Batalha}
\affiliation{Department of Astronomy and Astrophysics, University of California, Santa Cruz, CA 95064, USA}

\author[0000-0002-4207-6615]{James Kirk} 
\affiliation{Department of Physics, Imperial College London, Prince Consort Road, London SW7 2AZ, UK}

\author[0000-0003-3204-8183]{Mercedes L\'opez-Morales} 
\affiliation{Center for Astrophysics ${\rm \mid}$ Harvard {\rm \&} Smithsonian, 60 Garden St, Cambridge, MA 02138, USA}

\author[0000-0002-6721-3284]{Sarah E. Moran}
\affiliation{Department of Planetary Sciences and Lunar and Planetary Laboratory, University of Arizona, Tuscon, AZ, USA}

\author[0009-0008-2801-5040]{Johanna Teske} 
\affiliation{Earth and Planets Laboratory, Carnegie Institution for Science, Washington, DC 20015, USA}

\author[0000-0003-4328-3867]{Hannah R. Wakeford} 
\affiliation{School of Physics, University of Bristol, HH Wills Physics Laboratory, Tyndall Avenue, Bristol BS8 1TL, UK}

\author[0000-0002-0413-3308]{Nicholas F. Wogan}
\affiliation{NASA Ames Research Center, Moffett Field, CA 94035, USA}

\begin{abstract}

Planets between the sizes of Earth and Neptune are the most common in the Galaxy, bridging the gap between the terrestrial and giant planets in our Solar System. Now that we are firmly in the era of JWST, we can begin to measure, in more detail, the atmospheres of these ubiquitous planets to better understand their evolutionary trajectories. The two planets in the TOI-836 system are ideal candidates for such a study, as they fall on either side of the radius valley, allowing for direct comparisons of the present-day atmospheres of planets that formed in the same environment but had different ultimate end states. We present results from the JWST NIRSpec G395H transit observation of the larger and outer of the planets in this system, TOI-836c (2.587 R$_{\earth}$, 9.6 M$_{\earth}$, T$_{\rm eq}$$\sim$665 K). While we measure average 30-pixel binned precisions of $\sim$24ppm for NRS1 and $\sim$43ppm for NRS2 per spectral bin, we do find residual correlated noise in the data, which we attempt to correct using the JWST Engineering Database. We find a featureless transmission spectrum for this sub-Neptune planet, and are able to rule out atmospheric metallicities $<$175$\times$ Solar in the absence of aerosols at $\lesssim$1 millibar. We leverage microphysical models to determine that aerosols at such low pressures are physically plausible. The results presented herein represent the first observation from the COMPASS (Compositions of Mini-Planet Atmospheres for Statistical Study) JWST program, which also includes TOI-836b and will ultimately compare the presence and compositions of atmospheres for 12 super-Earths/sub-Neptunes.

\end{abstract}


\section{Introduction} \label{sec:intro} 
One of the most surprising revelations from the \textit{Kepler} mission was the ubiquity of super-Earths and sub-Neptunes, two populations of planets with sizes between that of Earth and Neptune (1--3.5 R$_\earth$) separated by a ``radius valley'' - a decrease in occurrence of planets with radii between 1.5 and 2.0 R$_\earth$ \citep{Batalha2013,Zeng2017, Fulton2018, VanEylen2018}. The larger of the two populations, sub-Neptunes (1.7 R$_\earth$ $\lesssim$ R$_{\rm p}$ $\lesssim$ 3.5 R$_\earth$; \citealt{Fulton2018}), are thought to host relatively large, primordial, hydrogen-dominated envelopes \citep{Venturini2020}, while the smaller super-Earths may possess higher mean molecular weight, ``secondary'' atmospheres outgassed from their interiors like those of the rocky planets in the Solar System \citep{Rogers2023}. How these two sub-classes of planets formed and evolved into the population we see today is a very active area of research. Various mechanisms have been suggested to sculpt a common population of planets into the two sub-classes. These mechanisms include photoevaporation \citep{Owen2013, Lopez2013}, core-powered mass loss \citep{Ginzburg2018,Gupta2019}, or a combination \citep[e.g.][]{Rogers2023b}. Alternatively, the division of the two populations may derive from their initial formation in a gas-rich versus gas-poor environment
 \citep[e.g.][]{Lee2022} or formation interior and exterior to the ice line \citep{luque2022, Burn2021}. Most recently, \cite{Burn2024} used a coupled formation and evolution model, including new equations of state and interior structure models in the treatment of water mixed with H/He. With this model, they showed that the radius valley can be interpreted as separating rocky, \textit{in-situ}-formed super-Earths from water-rich sub-Neptunes formed farther out in the disk. Each of these mechanisms can explain the overall radius distribution of the population, but specific mechanisms result in differing predictions about the atmospheric compositions of planets across this distribution (and with other variables like stellar mass and system age). Therefore, understanding the atmospheres of these small planets is crucial to understanding how they formed. 

For example, it may be that some sub-Neptunes did not form at their current locations, but outside of the snow line and then migrated inward, versus more rocky super-Earths that formed interior to the snow line \citep[e.g.,][]{Raymond2018, Mousis2020_iop,Burn2021}. Evidence for this scenario might be bulk planetary compositions that have a significant fraction of water/ice (``water worlds''). 
Recently, building on the findings of \cite{zeng2019} but using just the bulk densities of these planets, \cite{luque2022} indeed identified three distinct populations of small planets around M dwarfs -- rocky, water-rich, and gas-rich. The authors thus suggest that densities (and therefore the bulk compositions), not radii, delineate the observed small planet sub-populations. However, \cite{Rogers2023} were able to match the observed small planet densities with self-consistent, physically-motivated evolution models (both for photoevaporation and core-powered mass loss) that yield rocky core planets with a range of H\textsubscript{2}/He mass fractions that scale with planet mass. That is, they find that the mass-radius space is actually compositionally degenerate, and not able to conclusively provide evidence for a population of water worlds (see also \citealt{Valencia2007,Rogers&Seager2010}). However, determining the atmospheric compositions of sub-Neptunes would help break the degeneracies and provide more definite evidence for this population, further underlining the importance of understanding the atmospheres of these objects in gaining a full picture of their formation and evolution. 

Previous efforts to observe the atmospheres of super-Earths and sub-Neptunes using ground-based observatories and the Hubble Space Telescope (HST) yielded only a handful of targets with robust evidence of molecular features, including HD~97658b \citep{Guo2020}, HD~3167c \citep{Guilluy2021, Mikal-Evans2021}, K2-18b \citep{Benneke2019}, and 55~Cnc~e \citep{Tsiaras2016}. In comparison, observations of most other objects in this size class were consistent with a lack of features caused by clouds and hazes and/or a lack of an atmosphere altogether, e.g.,  GJ~1214b \citep{Kreidberg2014}, LHS~1140b \citep{Edwards2021}, L~98–59c and d \citep{Barclay2023,Zhou2023}, Kepler-138d \citep{Piaulet2022},  LHS 3844b \citep{diamondlowe2020}, LTT 1445Ab \citep{diamondlowe2023},  GJ 1132b \citep{libbyroberts2022}, and TRAPPIST-1~b, c, d, e, f, g, and h \citep{Wakeford2019,dewit2016,dewit2018,Garcia2022}.

More recently, JWST observations of sub-Neptunes ($\sim$2-4~R$_{\oplus}$) have started to reveal new insights into the compositions of these planets. Phase-curve observations of GJ 1214b (8.17 M$_{\earth}$, 2.628 R$_{\earth}$; \citealt{Cloutier2021,Gao2023}) indicated that while the planet had a featureless transmission spectrum with HST observations, at MIRI LRS wavelengths ($\sim$5-12$\mu$m), there is evidence of a molecular feature at $>$3$\sigma$, which has been attributed as most likely due to the presence of water vapor, on both the dayside and nightside of the planet \citep{Kempton2023, Gao2023}. 

Transmission observations with NIRISS SOSS and NIRSpec G395H of another sub-Neptune planet K2-18b (8.63 M$_{\earth}$, 2.61 R$_{\earth}$ ; \citealt{Cloutier2019}) detected CH$_4$ and CO$_2$ in an H$_2$-rich atmosphere \citep{Madhu2023}. \citet{Madhu2023} argued the carbon species and lack of detected NH$_3$ are consistent with a liquid water ocean under a H$_{2}$-rich atmosphere, suggesting that at least part of the radius valley may indeed be sculpted by the presence of these water worlds. However, more recently, \citealt{Wogan2024} used photochemical and climate models to show that the K2-18b data does not necessarily imply an ocean and can instead be explained by a $\sim 100\times$ Solar metallicity atmosphere sub-Neptune. In any case, it appears likely our knowledge surrounding sub-Neptune atmospheres will grow quickly with more JWST observations.

Up to present, JWST observations of super-Earths (1.0 R$_\earth$ $\lesssim$ R$_{\rm p}$ $\lesssim$ 1.7 R$_\earth$; \citealt{Fulton2018}) have been less informative than similar observations of sub-Neptunes. Using recent NIRSpec G395H transmission observations of the super-Earth GJ 486b (1.3 R$_{\earth}$, 3.0 M$_{\earth}$), \citetalias{Moran2023} found a slope in the spectrum consistent with either stellar activity or a water-rich atmosphere. In two different observations of GJ~1132b, \citetalias{May2023} found one of their observations is consistent with a water dominated atmosphere with $\sim$1\% methane, and the other observation is featureless. These ambiguous JWST observations of super-Earths demonstrate that further efforts are necessary to understand the smallest planets and their atmospheres or lack thereof, and the spectra of feature-less to feature-full 1--3.5 R$_\earth$ planets.

Motivated by the prevalence of 1-3~R$_{\earth}$ planets in general and the new possibilities for understanding their atmospheres unlocked by JWST, we initiated the large JWST Cycle 1 program COMPASS (Compositions of Mini-Planet Atmospheres for Statistical Study) (GO-2512, PIs N. E. Batalha \& J. Teske), a transmission spectroscopy survey of 11 super-Earth/sub-Neptune planets with NIRSpec G395H (our full sample also includes GTO target TOI-175.02 for a total of 12 planets; \citealt{Batalha2021}). Broadly, the program aims to better understand whether small planets have atmospheres, and if so, the compositional diversity of this population. More specifically, COMPASS aims to: (1) map out atmospheric detectability as a function of radius across the small planet regime\textemdash i.e. is there a radius below which we stop seeing detectable atmospheric features? (2) explore what the diversity in atmospheric composition implies for the origin(s) of these planets, (3) compare the compositions of ``sibling'' planets orbiting the same star but which have different radii/periods, and (4) investigate what population-level inferences we can draw from the entire sample. Importantly, the sample was selected with a quantitative, reproducible metric to enable robust inferences about the properties of somewhat cool ($\sim$400-1000 K), 1-3 R$_{\earth}$ planets. More details about the motivation and benefits of our selection method, and an example of a simple population-level trend which could be investigated, can be found in \cite{Batalha2023}.

Here we present observations of TOI-836c (TOI-836.01, HIP 73427c), a 2.587$\pm$0.088~R$_{\earth}$, 9.6$^{+2.7}_{-2.5}$ M$_{\earth}$ sub-Neptune \citep{Hawthorn2022} discovered by the Transiting Exoplanet Survey Satellite (TESS; \citealt{Ricker2015}) and confirmed by ground-based transit and radial velocity follow-up. The bulk density of the planet ($\sim$0.6 $\rho$$_{\earth}$) suggests a relatively extended gaseous envelope. TOI-836c is one of two planets known to exist in this system and exhibits transit timing variations (TTVs) on the order of 20 minutes. However, its sibling, TOI-836b, a super-Earth in an interior orbit, does not exhibit any TTVs, pointing to the possible existence of an additional exterior planet. While these planets were chosen agnostically using our aforementioned selection criteria \citep{Batalha2023}, the presence of the two confirmed planets in this multi-planet system in our COMPASS sample allows us to directly compare two planets on either side of the radius valley that formed within the same stellar environment. The transmission spectrum of the interior planet, TOI-836b, as well as a comparison of the NIRSpec G395H transmission spectra of the two confirmed planets in this system are presented in \citet{Alderson2024_836.02}.

In Sections~\ref{sec:observations} and~\ref{sec:reductions} we detail our observations of TOI-836c and present three different data reductions. In Section~\ref{sec:results} we present our fits to the white light curves and our transmission spectrum. In Section~\ref{sec:models} we interpret our transmission spectrum through the use of different atmospheric models. In Section~\ref{sec:discussion} we contextualize our results and present new bulk composition models for TOI-836c. We present our conclusions in Section~\ref{sec:conclusion}.

\begin{deluxetable}{ll}

\tablewidth{0pt}
\tablehead{\colhead{Property}&\colhead{Value}}
\startdata
K (mag)  & 6.804 $\pm$ 0.018 \\
T$_*$ (K)& 4552$\pm$154\\
R$_*$ (R$_\odot$)& 0.665$\pm$0.010\\
log(g) & 4.743$\pm$0.105\\
$[$Fe/H$]_{*}$& -0.284$\pm$-0.067\\
\hline
Period (days)& 8.59545 $\pm$  0.00001\\
Mass (M$_{\earth}$)&  9.6$^{+2.7}_{-2.5}$\\
Radius (R$_{\earth}$)& 2.587 $\pm$  0.088\\
T$_{\tt eq}$ (K)& 665$\pm$27\\
$e$& 0.078$\pm$0.056\\
$\omega$ (\textdegree)&-28 $\pm$ 113\\
inclination (\textdegree)&88.7 $\pm$1.5\\
a (AU)& 0.0750$\pm$0.0016
\tablecaption{System Properties for TOI-836c}
\enddata 
\label{table:system}
\tablenotetext{}{All values from \cite{Hawthorn2022}}
\end{deluxetable}

\section{Observations}\label{sec:observations}
One transit of TOI-836c was obtained by the Near-Infrared Spectrograph (NIRSpec) G395H/F290LP grating on 16 February 2023 in the NIRSpec Bright Object Timeseries mode with the SUB2048 subarray and the NRSRAPID readout pattern. We utilized 3 groups per integration, with 6755 integrations, and a total exposure time of 6.8 hours. The spectral traces using NIRSpec G395H (2.87 - 5.18 microns) are taken across two separate detectors, NRS1 and NRS2, with a small gap in wavelength coverage (between 3.72 and 3.82 microns) caused by the gap between the two detectors. 

\section{Data Reduction}\label{sec:reductions}
We reduced our observations using three independent pipelines to determine the robustness of our derived best-fit parameters and transmission spectrum. We detail our \texttt{Eureka!} \citep{Bell2022} reduction in Section~\ref{sec:eureka}, our \texttt{\texttt{ExoTiC-JEDI}} \citep{Alderson2022, Alderson2023} reduction in Section~\ref{sec:jedi}, and our \texttt{Aesop} (Alam et al. in prep) reduction in Section~\ref{sec:aesop}.

\subsection{\texttt{Eureka!} Reduction}\label{sec:eureka}
We first reduced the data using \texttt{Eureka!}\footnote{\url{https://github.com/kevin218/Eureka}} \citep{Bell2022}, an end-to-end pipeline for analyzing both HST and JWST data. The JWST \texttt{Eureka!} pipeline relies on the STScI \texttt{jwst} pipeline \citep{Bushouse2022} for its initial two stages (we use version 1.8.2 of the \texttt{jwst} pipeline and version 0.9 of \texttt{Eureka!}). We analyzed data from the two detectors separately in order to ascertain any flux calibration offsets between the detectors. Initial calibration was done in Stage 1 of \texttt{Eureka!}, following the suggested \texttt{jwst} pipeline steps of converting the raw data ramps to slopes. We utilize the default \texttt{Eureka!} parameters, with the exception of using a jump detection threshold of 15.  Additionally, we utilized group-level background subtraction to account for the 1/$f$ noise, which has been found to often be the dominant source of systematic noise in NIRSpec data \citep{Alderson2023, Rustamkulov2023}. We did not utilize the in-built \texttt{Eureka!} group level background subtraction method, instead opting to utilize the  1/f noise correction methodology employed in \texttt{ExoTiC-JEDI}\footnote{\url{https://github.com/Exo-TiC/ExoTiC-JEDI}} (see \cite{Alderson2023} for more details about the 1/f noise correction), as it provided more robust results with fewer temporal outliers. We used the default parameters for  \texttt{Eureka!} Stage 2, which again utilized the default reduction steps in the \texttt{jwst} pipeline in order to further calibrate the Stage 1 outputs. 

We then utilized Stage 3 of \texttt{Eureka!}, which does the data reduction and allows for different choices of trace and background apertures, different sigma thresholds for rejection of background pixels, and different polynomial orders for an additional column-by-column background subtraction. In addition to masking the pixels flagged as do not use by the \texttt{jwst} pipeline (which are masked by default within Stage 3 of \texttt{Eureka!}), we also mask pixels flagged as saturated, dead, hot, low quantum efficiency, or no gain value, as well as any individual columns that resulted in anomalous time series compared to the columns around it (there were nine such columns in NRS2). We then generated white light curves (summing over 2.862704--3.714356~$\mu$m for NRS1 and 3.819918--5.082485~$\mu$m for NRS2) using combinations of: extraction aperture half-widths of 4-8 pixels, background aperture half-widths of 8-11 pixels, sigma thresholds for outlier rejection for the optimal extraction of 10 and 60 (the latter of which approximates standard box extraction), and two different methods of background subtraction (an additional column-by-column background subtraction and a full frame median background subtraction). In order to determine the optimal combination of these extraction parameters, we generated a white light curve using Stage 4 of \texttt{Eureka!} for each combination, and chose the reduction that produced the lowest median absolute deviation in the white light curves. This process was done separately for NRS1 and NRS2, meaning that the reduction parameters that resulted in the optimal light curves were not necessarily the same for each detector. For NRS1, our optimal aperture half-width for the extraction of the trace was 4 pixels and our optimal inner bound for the background subtraction was 9 pixels (i.e. our background was from 9 pixels away from the trace location outwards to the lower and upper edges of the frame). For NRS2, our optimal aperture half-width for the extraction of the trace was 6 pixels and our optimal inner bound for the background subtraction was 8 pixels. For both detectors, we favored an additional column-by-column background subtraction over a median frame background subtraction and a sigma threshold for the optimal extraction of 60. We then utilized a custom light curve fitting code instead of the default \texttt{Eureka!} fitting code for increased flexibility, but here continue to refer to this reduction as the \texttt{Eureka!} reduction.

\begin{figure*}
\begin{centering}
\includegraphics[width=\textwidth]{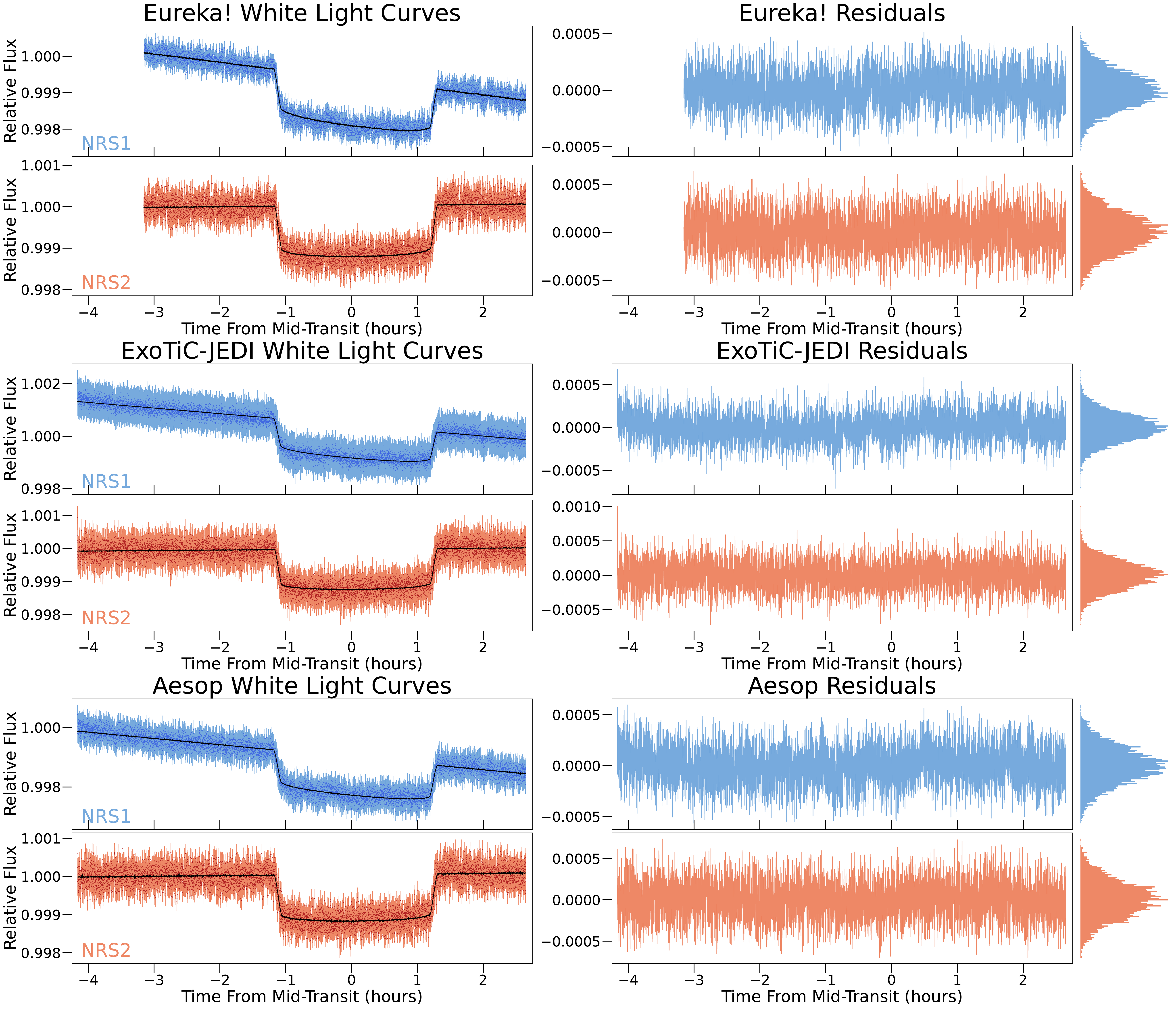}
\caption{White light curves for each reduction with the best-fit models and associated residuals shown as both a time series and histogram. NRS1 (blue) and NRS2 (red) are shown separately for each reduction. The \texttt{Eureka!} reduction is shown at the top, \texttt{ExoTiC-Jedi} in the middle, and \texttt{Aesop} at the bottom. }
\label{figure:WLC}   
\end{centering}
\end{figure*}

From our white light curves, we then produced spectroscopically binned light curves of 30-pixels (R$\sim$200). We iteratively trim 3$\sigma$ outliers three times from a 50 point rolling median in both the white light curves and the binned light curves. We then fit the optimal light curves for NRS1 and NRS2 separately using a combination of a Levenberg–Marquardt least-squares minimization and the affine-invariant Markov chain Monte Carlo (MCMC) ensemble sampler \texttt{emcee} \citep{Foreman-Mackey2013} where we fit for a combined astrophysical and systematic noise model and optimize the log-likelihood. Our astrophysical model is calculated using the \texttt{batman} package \citep{Mandel2002,Kreidberg2015}, assuming quadratic limb-darkening coefficients calculated using ExoTiC-LD \citep{Grant2022} using Set One of the MPS-ATLAS models \citep{Kostogryz2022,Kostogryz2023}. We fit for the system inclination, a/R${_*}$ (the orbital semi-major axis with respect to the host star), T$_0$ (time of transit), and R$_{p}$/R$_{*}$ (the radius of the planet with respect to the radius of the host star), fixing the period,  eccentricity, and stellar parameters to those shown in Table~\ref{table:system}. Our instrumental noise model, \textit{S}, was of the form
\begin{equation}
S= p_{1} + p_{2}*T+ p_{3}*X + p_{4}*Y , 
\label{eq:1}
\end{equation}
where $p_{N}$ is a free parameter, $T$ is the vector of times, and $X$ and $Y$ are vectors of the positions of the trace from Stage 3 of \texttt{Eureka!}. We also fit for an additional error term that we add in quadrature with the per-point measured errors. We trim the initial 1,000 points to remove any initial ramp.  We chose to trim a conservatively long initial duration to account for any initial ramp in our data that might bias the linear slope in our systematic noise model given our abundance of pre-transit baseline and the apparent small initial ramp in the NRS1 white light curve (see Figure \ref{figure:WLC}). We initialized three times the number of free parameters for the number of walkers (for a total of 27 walkers for these fits) at the best-fit values from the initial Levenberg–Marquardt least-squares minimization. For the MCMC fit, we used uninformed priors for all of our parameters and used a burn-in of 50,000 steps which was discarded then an additional 50,000 steps to ensure adequate sampling of the posterior. We adopt the median of each chain as our best-fitting value for each parameter and take the standard deviation of each chain as the uncertainty for each parameter. We show the white light curves for NRS1 and NRS2, the best-fit models, and the residuals in the top panels in Figure~\ref{figure:WLC} and the best-fit astrophysical parameters in Table~\ref{tab:bestfit}.

We used the resulting MCMC chains of the astrophysical parameters from the white light curve fits as Gaussian priors on the spectroscopic light curve fits. We combined the MCMC chains from the NRS1 and NRS2 fits for the inclination, a/R$_{*}$, and T$_{0}$ and used the resulting combined chains as Gaussian priors for each of those parameters for the fits to the spectroscopically binned light curves centered at the median of each of the combined chains and with a conservative width of 3$\times$ the standard deviation of the combined chains. We utilized a wide flat prior for the R$_{p}$/R$_{*}$ of each spectroscopic bin.

In order to further validate the robustness of our results, we then reduced our data with additional independent reductions, using \texttt{ExoTiC-JEDI} and {\tt Aesop}.

\subsection{\texttt{ExoTiC-JEDI} Reduction}\label{sec:jedi}

The Exoplanet Timeseries Characterisation - JWST Extraction and Diagnostic Investigator (\texttt{ExoTiC-JEDI}, \citealt{Alderson2022}) package produces an end-to-end reduction of JWST time-series data, beginning with the \texttt{uncal} files. Throughout the analysis, data from NRS1 and NRS2 are treated separately. 

\texttt{ExoTiC-JEDI} begins with a modified version of Stage 1 of the \texttt{jwst} pipeline (v.1.8.6, context map 1078), performing linearity, dark current and saturation corrections, and using a jump detection threshold of 15. Next, a custom destriping routine is used to remove 1/$f$ noise, masking the spectral trace at 15 times the standard deviation of the PSF from the dispersion axis in each integration, subtracting the median pixel value from each detector column at the group level. We also perform a custom bias subtraction, computing the median of each detector pixel in the first group across all the integrations in the time series, and then subtracting this new median ``pseudo-bias'' image from all groups. This method was found to improve the standard deviation of the out-of-transit points for both detector white light curves, and reduced an offset seen between the resulting NRS1 and NRS2 transmission spectra. \texttt{ExoTiC-JEDI} uses the standard \texttt{jwst} Stage 1 ramp fitting step and Stage 2 steps to produce the 2D wavelength map.

We next extract our 1D stellar spectra, with further pixel cleaning steps and 1/$f$ correction. With the standard data quality flags produced by the \texttt{jwst} pipeline, we replace any pixels flagged as do not use, saturated, dead, hot, low quantum efficiency, or no gain value, with the median value of the next 4 pixels in each row. We additionally identify any remaining pixels that are outliers from their nearest neighbors on the detector, or throughout the time series (such as cosmic rays), with a 20$\sigma$ threshold in time and a 6$\sigma$ threshold spatially. The flagged pixel is replaced with the median value of the surrounding 10 integrations or 20 pixels in the row. Remaining 1/$f$ noise and background is removed by subtracting the median unilluminated pixel value from each column in each integration. To extract the 1D stellar spectra, we fit a Gaussian to each column to obtain the spectral trace and width, fitting a fourth-order polynomial to each and smoothing with a median filter to determine the aperture region. We used an aperture five times the FWHM of the trace, approximately 8 pixels wide, using an intrapixel extraction. The resulting 1D stellar spectra are cross-correlated to obtain the $x$- and $y$-positional shifts throughout the observation for use in systematic light curve detrending.

We fit white light curves for both NRS1 and NRS2, as well as spectroscopic light curves across the full NIRSpec/G395H wavelength range. For the white light curves (spanning 2.814--3.717\,$\micron$ for NRS1 and 3.824--5.111\,$\micron$ for NRS2), we fit for $i$, a/R$_{*}$, T$_0$, and R$_{\rm p}$/R$_{*}$ (Table~\ref{tab:bestfit}), holding the period and eccentricity fixed to values presented in \citet{Hawthorn2022} (see Table~\ref{table:system}). We use stellar limb darkening coefficients calculated with \texttt{ExoTiC-LD} based on the stellar T$_{*}$, log(g), and [Fe/H]$_{*}$ listed in Table \ref{table:system}, using Set One of the MPS-ATLAS stellar models \citep{Kostogryz2022, Kostogryz2023} and the non-linear limb darkening law \citep{Claret2000}, which are held fixed in a \texttt{batman} transit model. We used a least-squares optimizer to fit the transit model simultaneously with our systematic model $S(\lambda)$, which took the form

\begin{equation}
S(\lambda) = s_0 + (s_1 \times x_{s}|y_{s}|) + (s_2 \times t) \mathrm{,}
\label{eq:EJ}
\end{equation}

\noindent where $x_s$ is the $x$-positional shift of the spectral trace, $|y_s|$ is the absolute magnitude of the $y$-positional shift of the spectral trace, $t$ is the time and $s_0, s_1, s_2$ are coefficient terms. We use the same binning scheme as was utilized for the \texttt{Eureka!} reduction. For these spectroscopic light curves, we fit for $R_p/R_*$, holding $T_0$, $i$, and $a/R_*$ fixed to the respective white light curve fit value. For all fitted parameters, we take the optimized values and the standard deviation errors as the value and uncertainty for each parameter. For both the white and spectroscopic light curves, we removed any data points that were greater than 4$\sigma$ outliers in the residuals, and refit the light curves until no such points remained. We also rescaled the flux time series errors using the beta value \citep{Pont2006} measured from the white and red noise values calculated for the Allan variance plots to account for any remaining red noise in the data (see middle panel of Figure~\ref{figure:WLC_RMS}).

\subsection{\texttt{Aesop} Reduction}\label{sec:aesop}

{\tt Aesop} is a JWST data reduction pipeline developed for the reduction and analysis of NIRSpec G395H observations that has been benchmarked with several other pipelines in the literature (see e.g., \citealt{Alderson2023}). The pipeline's methodology is described in detail in Alam et al. (in prep), which we briefly summarize here. We began with the {\tt uncal} uncalibrated JWST data products, treating the data sets for the NRS1 and NRS2 detectors separately. We first ran the standard Stage 1 steps of the {\tt jwst} pipeline for time-series observations, including corrections for saturation, bias, linearity, and dark current. We set the detection threshold for the jump step to 15, apply a custom group-level background subtraction to remove 1/$f$ noise using a 15$\sigma$ threshold and a second-order polynomial, and perform standard ramp fitting. We used the standard steps of the {\tt jwst} pipeline for Stage 2 to extract the integration exposure times and the 2D wavelength array. 

We then performed additional cleaning steps to replace poor data quality pixels (flagged as bad, saturated, hot, or dead) with the median of the neighboring pixels and remove residual 1/$f$ noise by calculating the median pixel value of each column. To extract the 1D time-series stellar spectra, we fit a Gaussian profile to each column of a given integration to find the center of the spectral trace and smoothed the trace centers with a median filter, followed by fitting a fourth-order polynomial to the smoothed trace centers. We then extracted the 1D stellar spectra by summing up the flux within a 10-pixel wide aperture, calculating the uncertainties in the stellar spectra assuming photon noise. 

We generated the white light curves for NRS1 and NRS2 by summing the flux between 2.862704--3.714356\,$\mu$m and 3.819918--5.082485\,$\mu$m, respectively, and inflated the flux uncertainties following \citet{Pont2006}. We then fit the broadband and spectroscopic light curves using a least-squares minimizer. We fit each light curve with a two-component function consisting of a transit model (generated using {\tt batman}) multiplied by a systematics model including a linear polynomial in the x- and y-pixel positions on the detector. We first fit the broadband light curve by fixing the period and eccentricity to the \citet{Hawthorn2022} values, and fitting for $T_{0}$, $a/R_{*}$, $i$, $R_{p}/R_{*}$, stellar baseline flux, and systematic trends using wide uniform bounds. We adopt the optimized parameters and their standard deviations as the best-fitting value and associated uncertainty for our fitted parameters. We again use the same binning scheme as was utilized for the other two reductions. For these spectroscopic light curves, we fixed $T_{0}$, $a/R_{*}$, and $i$ to the best-fit values from the broadband light curve (Table~\ref{tab:bestfit}) and fit for $R_{p}/R_{*}$. We held the non-linear limb-darkening coefficients fixed to theoretical values, which we computed using {\tt ExoTiC-LD} and the 3D stellar model grid from \citet{Magic2015}.    

\section{Results}\label{sec:results}
\subsection{Correlated noise and timing offsets}\label{sec:engineering}
\begin{centering}
\begin{figure}[ht!]
\includegraphics[height=.45\textwidth]{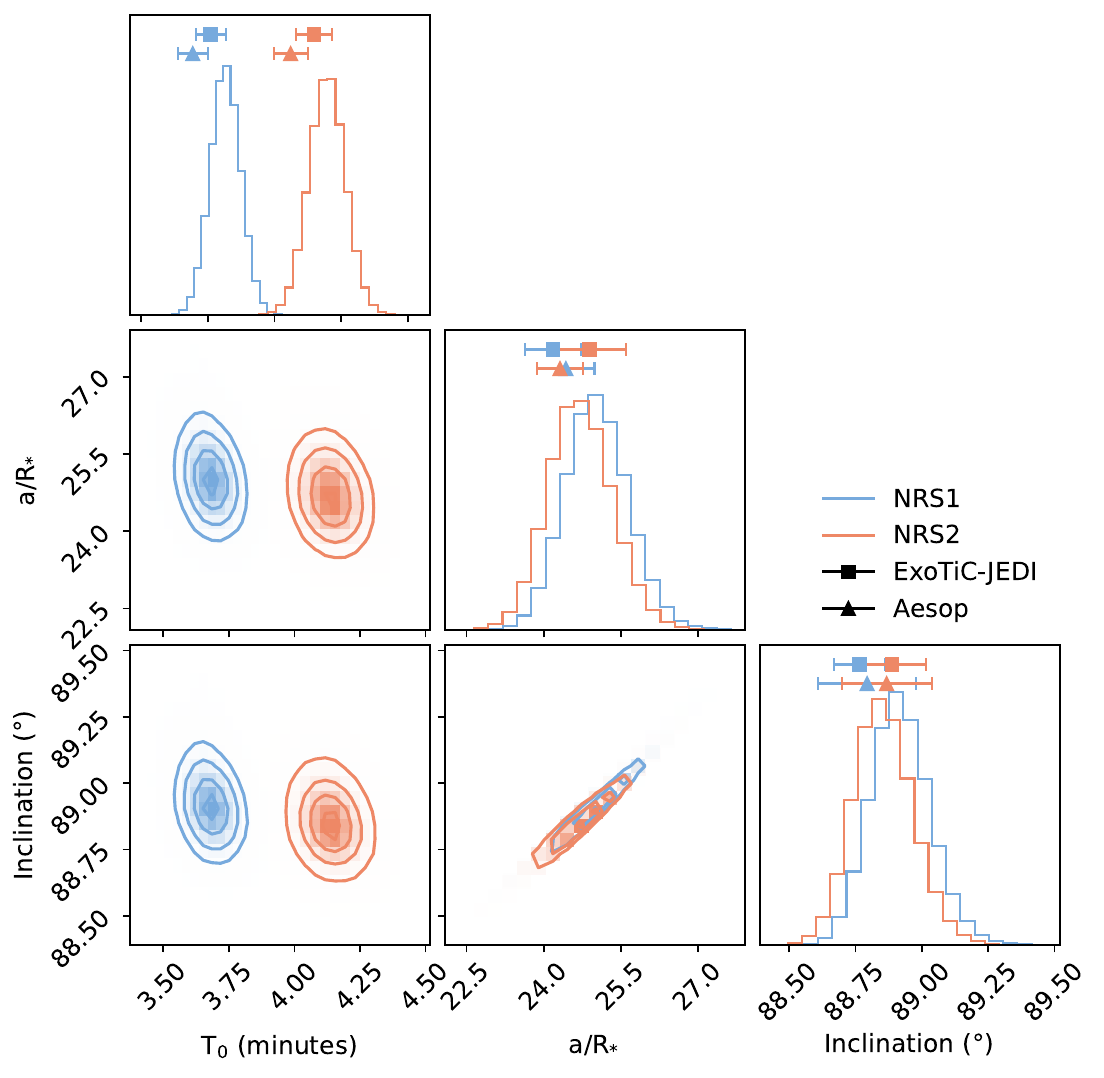}
\caption{We show the time in minutes from the predicted mid-transit time of 59991.7252 BMJD (T$_{0}$), the a/R$_{*}$, and the inclination from the fits to our white light curves. The contours show the posterior distributions from the MCMC fits to the \texttt{Eureka!} reduction and the best-fit values from the fits to the other reductions are shown as points above the distribution for each parameter (as we did not run an MCMC for the other reductions). All of our reductions agree on the presence of an offset in the times of transit between NRS1 and NRS2.}
\label{figure:corner}   
\end{figure} 
\end{centering}

\begin{figure*}
\begin{centering}
\includegraphics[width=\textwidth]{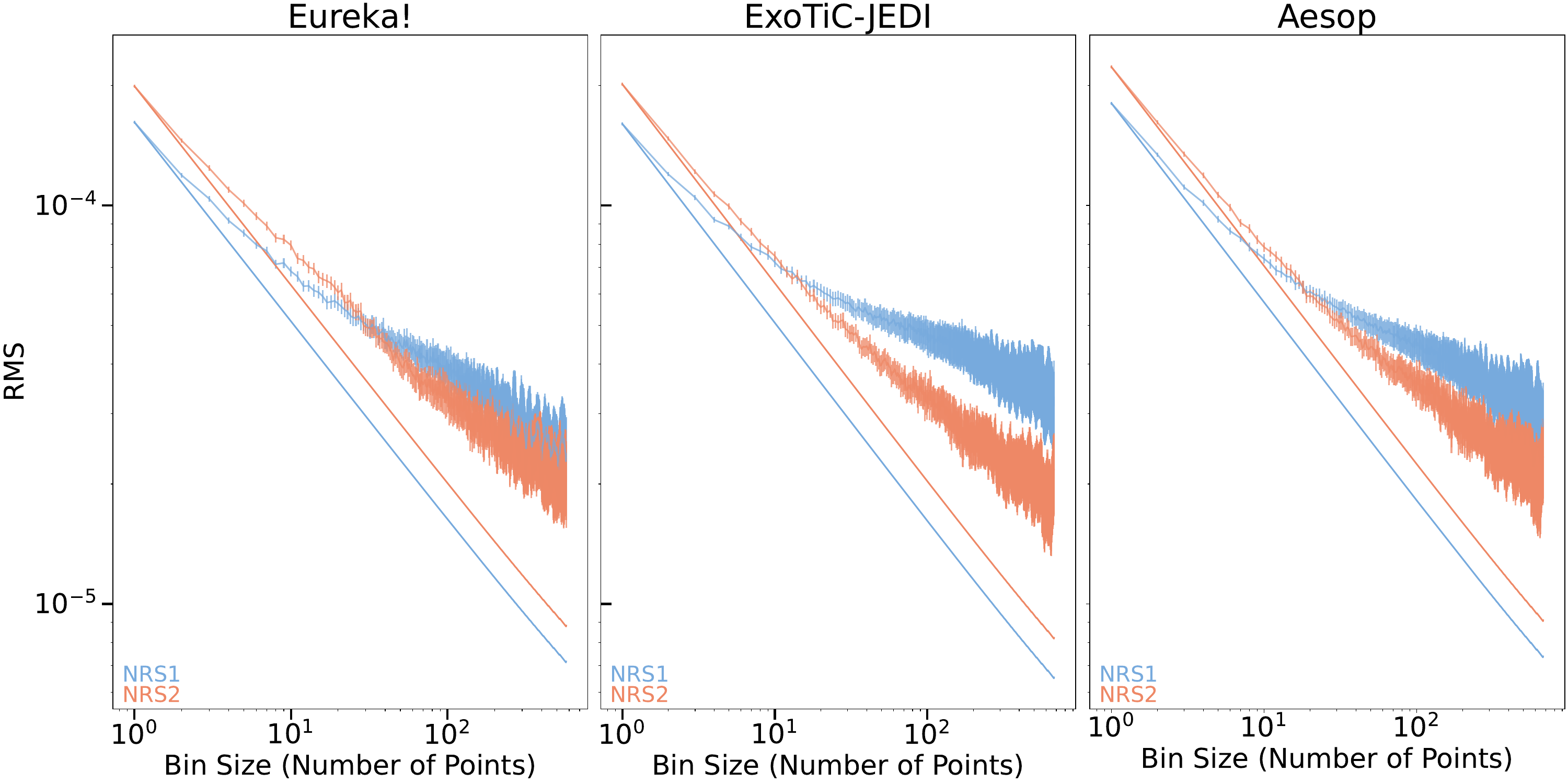}
\caption{The Allan variance plots for both detectors for each reduction are shown. We show the actual noise and expected noise for NRS1 in blue and for NRS2 in red. In the absence of red noise, the residuals would follow the solid lines. We see residual correlated noise in both detectors for all reductions, but especially for NRS1.} 
\label{figure:WLC_RMS}   
\end{centering}
\end{figure*}

\begin{figure}
\begin{centering}
\includegraphics[width=.5\textwidth]{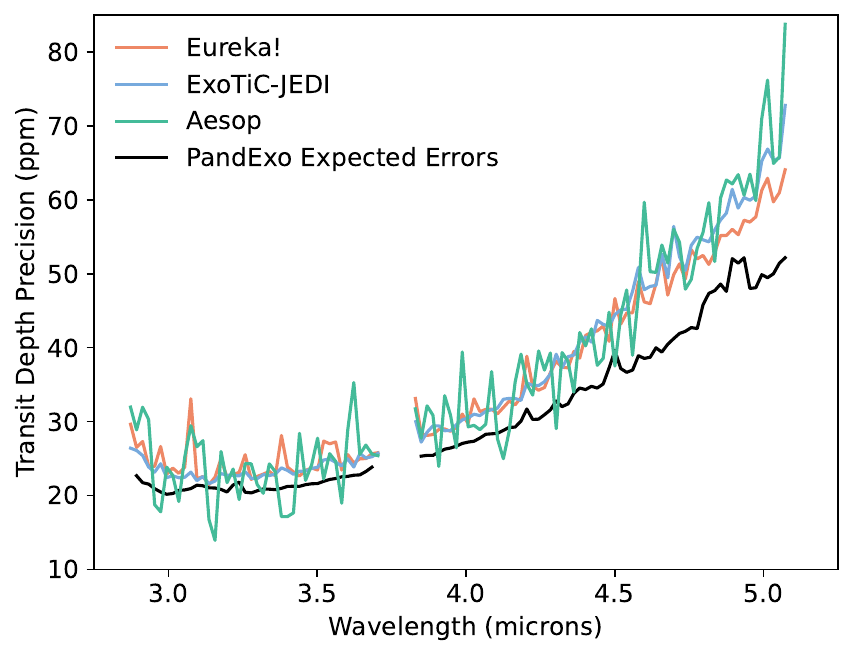}
\caption{The expected errors on our transmission spectrum simulated from \texttt{PandExo}  \citep{Batalha2017} with our measured errors for each of our three reductions. We achieve comparable precisions between our three reductions, and generally are not able to achieve errors as expected from \texttt{PandExo}, likely due to the noise structure in our observation that only utilizes three groups.} 
\label{figure:errors}   
\end{centering}
\end{figure}

The fits to the white light curves for NRS1 and NRS2 for all three independent reductions were consistent in their best-fit parameters (see Table~\ref{tab:bestfit}). For all three reductions, the best-fit times of mid-transit differed by greater than 4$\sigma$ between the two detectors (see Figure~\ref{figure:corner} and Table~\ref{tab:bestfit}). This is likely not a physical effect, as the NRS1 and NRS2 data comprise a single observation in which the light was dispersed across the two detectors simultaneously. While the detectors are utilized simultaneously, the time stamps between the two detectors vary slightly. It is important to note that this time array difference is more than an order of magnitude smaller than the timing offset between the transit times from the fits to the white light curves of the two detectors, so is not likely to be the reason for the transit timing offset. Due to the fact that all three reductions resulted in the same timing offset despite different reduction parameter choices and fitting methods, this offset is likely also not reduction or fitting specific, but instead due to the residual correlated noise present in our fits to the white light curves (Figure~\ref{figure:WLC_RMS}). Using the reduction and fitting method described in Section \ref{sec:eureka}, we also investigated whether utilizing a higher order and more complicated systematic noise model could reduce the amount of correlated noise and the timing offset, but did not find that the added complexity reduced the amount of residual correlated noise or remedied the timing offset. We note that the precisions we achieve, despite the correlated noise, are still comparable to those expected from simulations of the observations from \texttt{PandExo} (\citealt{Batalha2017}, see Figure~\ref{figure:errors}). 

\begin{deluxetable*}{llll}
\tablewidth{0pt}
\tablehead{\colhead{} &\colhead{\texttt{Eureka!}}&\colhead{\texttt{ExoTiC-JEDI}}&\colhead{\texttt{Aesop}}}
\startdata
NRS1 T$_0$ (days)\tablenotemark{a} &0.002555 $\pm$ 4.7$\times$10$^{-5}$ & 0.002510 $\pm$ 4.7$\times$10$^{-5}$ & 0.002453 $\pm$ 4.7$\times$10$^{-5}$  \\
NRS1 R$_{\rm p}$/R$_{*}$&0.035041 $\pm$ 9.3 $\times$10$^{-5}$ & 0.034841  $\pm$ 9.3$\times$10$^{-5}$ & 0.034957 $\pm$ 8.3$\times$10$^{-5}$  \\
NRS1 a/R$_{*}$ &25.01 $\pm$ 0.62 & 24.17 $\pm$ 0.54 & 24.43 $\pm$ 0.56 \\
NRS1 inclination (\textdegree)& 88.92 $\pm$ 0.11 & 88.76 $\pm$ 0.10 & 88.79 $\pm$ 0.18 \\
NRS2 T$_0$ (days)\tablenotemark{a} &0.002873 $\pm$ 5.7$\times$10$^{-5}$ & 0.002833$\pm$ 5.8$\times$10$^{-5}$ & 0.002759 $\pm$ 5.3$\times$10$^{-5}$  \\
NRS2 R$_{\rm p}$/R$_{*}$&0.034317 $\pm$ 9.7$\times$10$^{-5}$ & 0.03425 $\pm$ 1.0 $\times$10$^{-4}$ &  0.034273 $\pm$ 9.8$\times$10$^{-5}$  \\
NRS2 a/R$_{*}$ &24.69 $\pm$ 0.62 & 24.87 $\pm$ 0.72 & 24.31 $\pm$ 0.45 \\
NRS2 inclination (\textdegree)& 88.85 $\pm$ 0.11 & 88.89 $\pm$ 0.13 & 88.87 $\pm$ 0.17
\tablecaption{Best-fit system parameters from the fits to the white light curves for all three reductions.}
\label{tab:bestfit}
\enddata 
\tablenotetext{a}{Time from expected mid-transit of 59991.7252 BMJD.}
\end{deluxetable*}

It is important to determine whether the correlated noise that is likely causing the offset in the transit times is affecting the resulting transit depths and therefore the overall transmission spectrum. We are also interested in pinning down the correct time of transit to allow for an improved understanding of the dynamics of this system. Moreover, given the evidence of TTVs on the order of 20 minutes for this planet \citep{Hawthorn2022}, we cannot adopt a fixed center of transit time and apply it to both the NRS1 and NRS2 white light curves since we cannot be certain which of the two derived T$_{0}$ values is closest to the truth. 

\begin{figure}
\begin{centering}
\includegraphics[width=.45\textwidth]{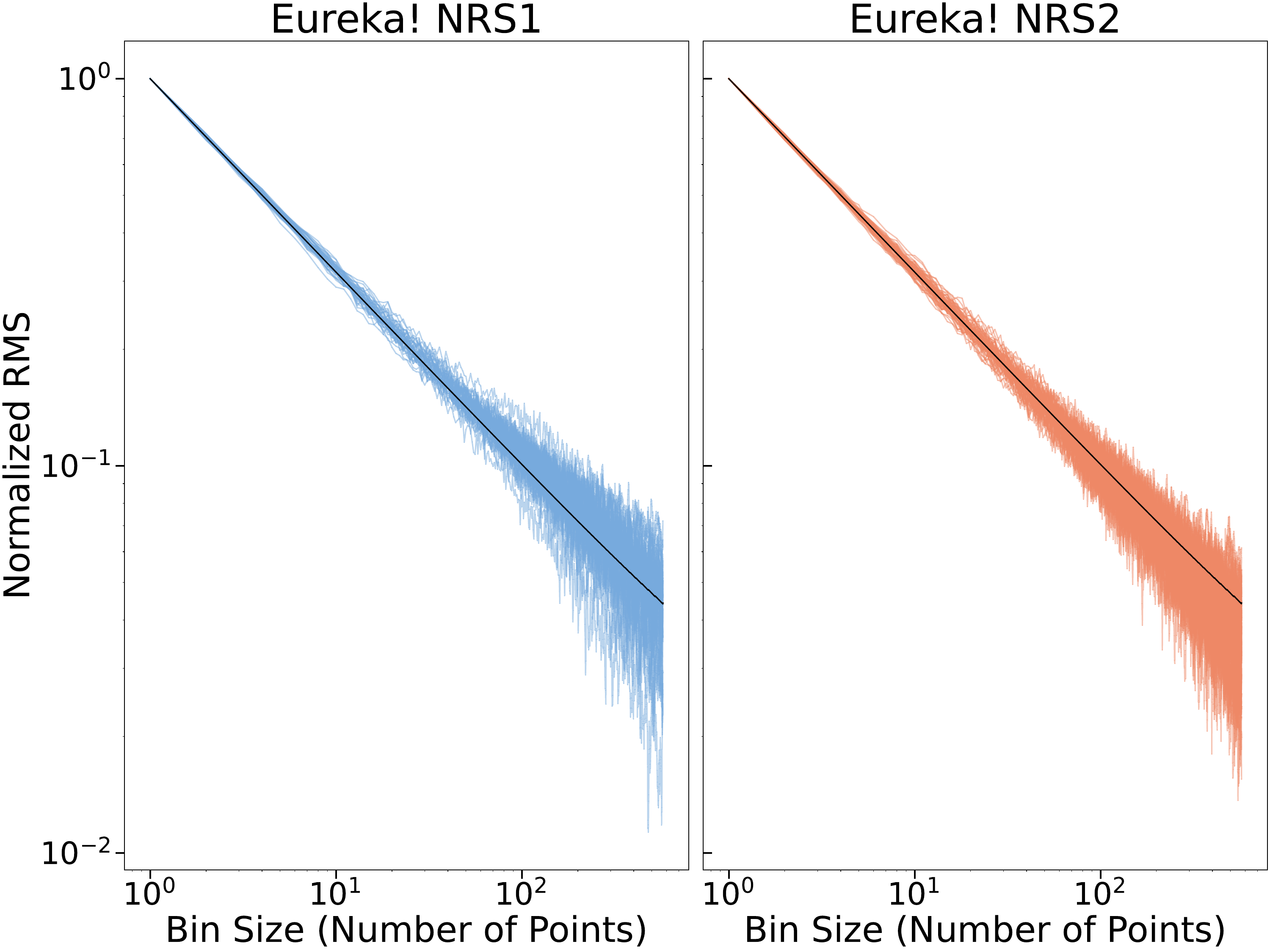}
\caption{The Allan variance plots for both detectors for the \texttt{Eureka!} reduction for each of the spectroscopic bins fit with a systematic model of the form in Equation~\ref{eq:1}. Each colored line represents a different bin and the black line represents the normalized standard error. We have normalized the root mean squared (RMS) values for the data in order to show that the bins follow a 1/$\sqrt{N}$ scaling and that there is little excess red noise in the residuals of the fits to the binned light curves.} 
\label{figure:bins_rms}   
\end{centering}
\end{figure} 

In order to investigate the cause of the residual correlated noise, we utilized the JWST Engineering Database. There are thousands of telemetry points, designated by mnemonics, that are taken onboard the spacecraft at different time intervals, which are all available via the Calibrated Engineering Data Portal\footnote{\url{https://mast.stsci.edu/portal/Mashup/Clients/jwstedb/jwstedb.html}}. We take all available mnemonics for both the spacecraft and the NIRSpec instrument corresponding to the time spanned by our observation of TOI-836c in order to determine whether decorrelating against these engineering parameters would decrease the correlated noise and reduce the offset in the transit times between the detectors.

We take the more than 15,000 NIRSpec and spacecraft associated mnemonics and do an initial cut, only selecting those that varied during our observation and represented time series that could be decorrelated against. We then took the resulting more than 1,000 mnemonics and resampled them to the NRS1 and NRS2 time arrays.  

We take an agnostic approach to determining which engineering parameters are part of our systematic model. We first individually add each engineering parameter to the systematic model in Equation~\ref{eq:1}, meaning that our new systematic model is of the form

\begin{equation}
S= p_{1} + p_{2}*T+ p_{3}*X + p_{4}*Y + p_{5}*E, 
\label{eq:3}
\end{equation}

\noindent where E represents the time series of a single engineering parameter that has been resampled to the time series for the white light curve from the detector that is being fit. We loop over each of the more than 1,000 engineering parameters separately, fitting using a Levenberg–Marquardt least-squares minimization. We use the same process and leave the same parameters as free as described in the fits in Section \ref{sec:eureka}, but elect to not run an MCMC for every combination for computational efficiency. 

We then generate a ranked list of all of the engineering parameters for each detector (as the engineering parameter exploration is done independently). We calculate the RMS of the difference between the measured standard deviation of the residuals and the predicted photon noise limit (i.e., the RMS of the difference of the expected and measured lines in Figure~\ref{figure:WLC_RMS}), and then rank all of the engineering parameters by this calculated metric. Ranking the parameters in this way ensures that their inclusion will decrease the amount of residual correlated noise in the white light curves.

In order to not be limited to a single additional engineering parameter (and indeed some of the mnemonics are associated with other mnemonics), we then combine the top ranked mnemonics. Treating each detector separately, we begin with only including the single mnemonic that improves the residual correlated noise the most (e.g. the top mnemonic in our ranked list described above) and fit the white light curve using the MCMC procedure detailed in Section \ref{sec:eureka}. We then add the second best mnemonic to our systematic noise model in the form 

\begin{equation}
S= p_{1} + p_{2}*T+ p_{3}*X + p_{4}*Y + p_{5}*E_{1} + p_{6}*E_{2}, 
\label{eq:4}
\end{equation}

where E$_{1}$ is the top ranked mnemonic and E$_{2}$ is the second highest ranked mnemonic. We then repeat the full MCMC fit on this two mnemonic engineering parameter model. We then add another mnemonic to the systematic model and run an MCMC fit. We continue this process to include up to 15 mnemonics, meaning that we have 15 different systematic noise models that include engineering parameters that we try for each detector. We then determine the optimal number of engineering parameters as the number that corresponds to the engineering model that most reduced the residual correlated noise without adding additional scatter. 

The purpose of this exercise was to determine if a reduction in residual correlated noise would result in times of transit that were more in agreement between the two detectors, and therefore if the transit timing offset was likely caused by the residual correlated noise. However, we are not claiming to determine an underlying physical process that is being tracked by our best engineering parameter model. We first optimized the two detectors separately, so it was not a requirement that the mnemonics utilized in the best engineering model be identical between the two detectors, or have the same number of additional engineering parameters in the systematic model. Interestingly, the two top mnemonics between the two detectors, IGDP\_MIR\_IC\_V\_DETHTR and IFGS\_TFG\_BADCNTCNT respectively, were in agreement. The former mnemonic is a focal plane electronic SCE1-IC Detector Heater Voltage and the latter is the Fine Guidance Sensor Track/Fine Guide Image Data Processing Telemetry Bad Centroid Data Count. While utilizing only these two mnemonics did improve the residual correlated noise, this alone did not remedy the transit timing offset, so we do not further discuss the physical implications of these parameters. The other top engineering parameters did not agree between the two detectors. 

For NRS2, we found that, in all systematic noise models including up to 15 different engineering terms, all of the times of transit were in agreement with each other (and the time derived from the non-engineering model) within 0.65$\sigma$. Thus, adding the engineering parameters did not make a significant difference in the mid-transit time for NRS2, although the inclusion of additional terms did help to reduce the amount of correlated noise. The amount of residual correlated noise in NRS2 is lower than for NRS1 regardless of the systematic model used and all of the inclinations and a/R$_{*}$ values in the fits with and without any number of engineering parameters were all within 0.5$\sigma$ of each other, meaning that while the additional engineering parameters helped to reduce the amount of residual correlated noise, any residual correlated noise from the fits using the nominal systematic model do not seem to be altering the NRS2 best-fit parameters. The NRS1 detector, on the other hand, shows much more residual correlated noise and the additional engineering terms more appreciably affect the red noise. Moreover, the addition of engineering terms results in times of transit that differ by up to 1.8$\sigma$ from the fit without the engineering parameters. 

For consistency, we opt to use the optimal engineering parameter systematic noise model (including 13 parameters) from NRS1 to fit both detectors. We find that the best-fit white light curve astrophysical parameters for NRS2 are still within 1$\sigma$ of those derived from the model without the inclusion of engineering parameters, further showing that any residual correlated noise in NRS2 is not altering the white light curve best-fit parameters. As different engineering models do not make a difference to our derived best-fit values for NRS2, we opt for this model (where we only use the NRS1 derived parameters) as our engineering model for simplicity.

\begin{figure}
\begin{centering}
\includegraphics[width=.45\textwidth]{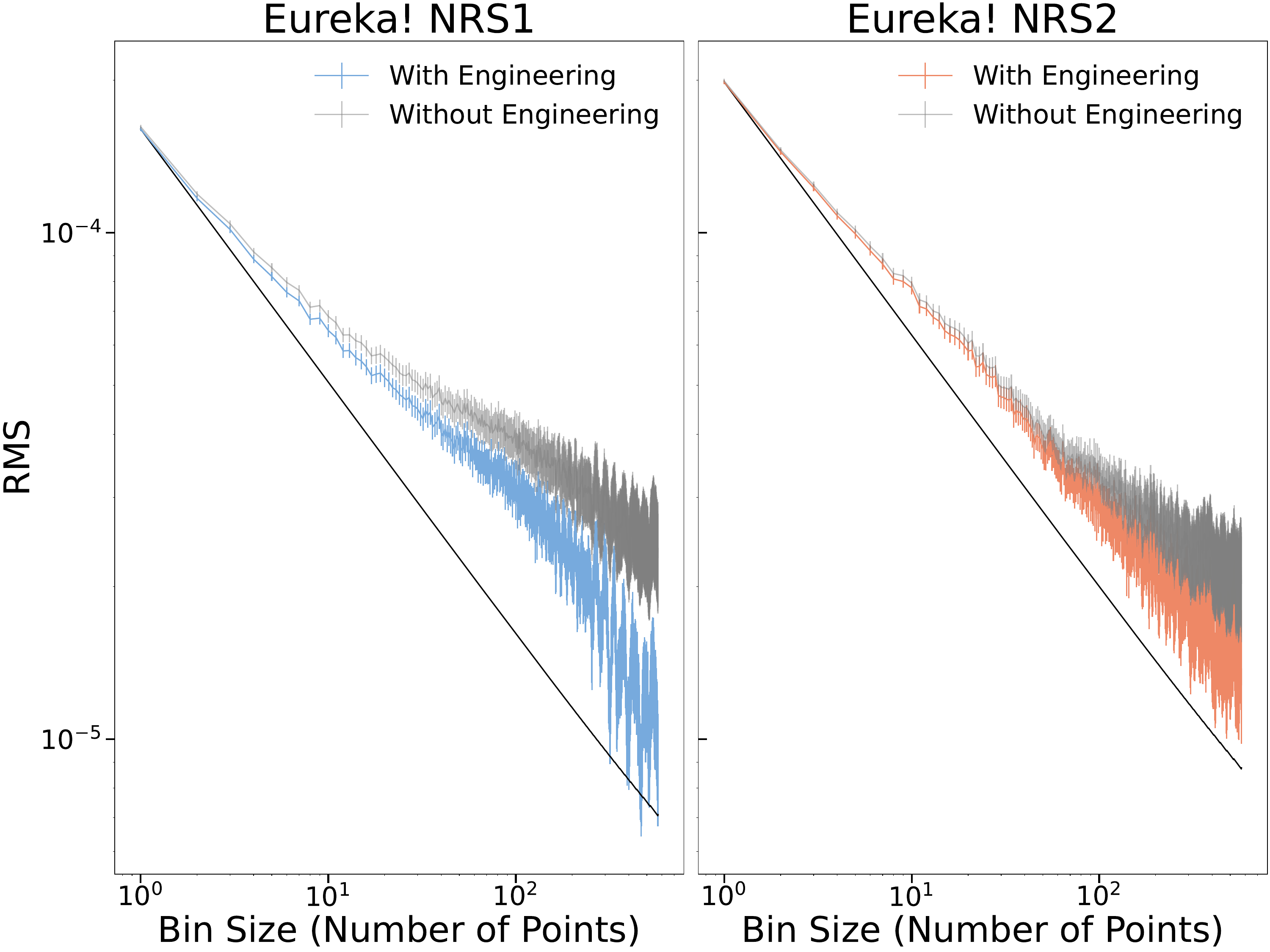}
\caption{The Allan variance plots for both detectors for the \texttt{Eureka!} reductions where we include 13 engineering parameters in the systematic model are shown in colors while the corresponding curves without the engineering parameters are shown in grey for each detector (the grey curves are identical to those shown in panel one of Figure~\ref{figure:WLC_RMS}). In the absence of red noise, the residuals would follow the black lines. While there is still residual correlated noise, the amount of red noise has been reduced.} 
\label{figure:RMS_engineering}   
\end{centering}
\end{figure}

\begin{centering}
\begin{figure}[ht!]
\includegraphics[height=.45\textwidth]{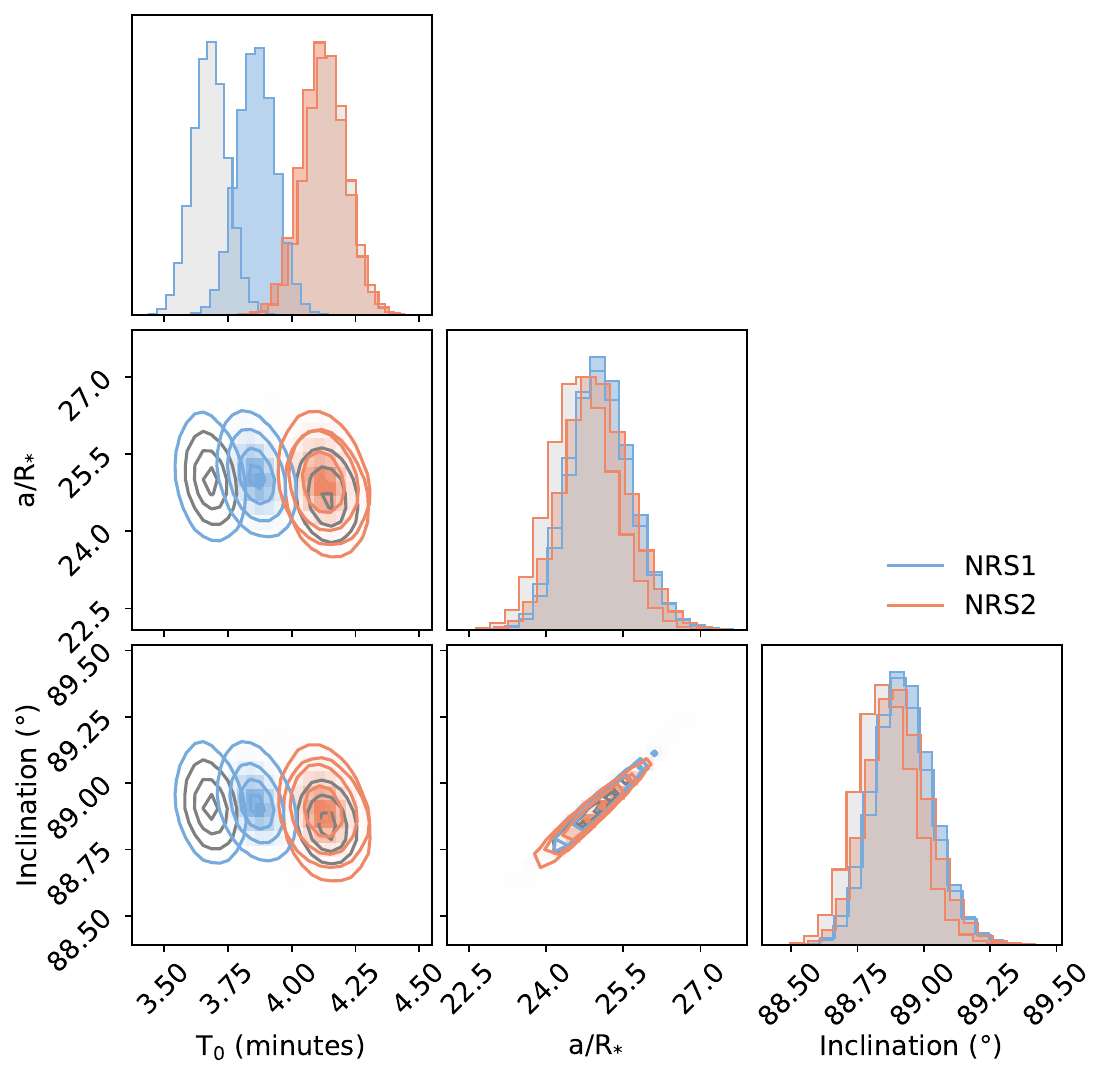}
\caption{The posteriors for the time (T$_{0}$ in minutes from the predicted mid-transit time of 59991.7252 BMJD), the a/R$_{*}$, and the inclination from the fits to our white light curves of our \texttt{Eureka!} reduction including 13 engineering parameters in our systematic model are shown in colors. The corresponding posteriors from the fits without the engineering parameters are shown in grey and outlined in their respective detector colors. The posteriors for the fits without the engineering parameters are identical to those shown in Figure~\ref{figure:corner}. The T$_{0}$ values are more in agreement between NRS1 and NRS2 when we include engineering parameters. }
\label{figure:corner_engineering}   
\end{figure} 
\end{centering}

We find that utilizing the best-fit engineering mnemonics from NRS1 for the systematic models for both NRS1 and NRS2 reduces the amount of residual correlated noise for both detectors as evident in Figure~\ref{figure:RMS_engineering} (although the inclusion of the engineering terms reduces the amount of residual correlated noise more for NRS1 than for NRS2), and causes the best-fitting times of transit to be within 2.3$\sigma$ of each other (Figure~\ref{figure:corner_engineering}). This means that while none of our 15 different systematic models that included engineering parameters could entirely resolve the discrepancy in transit times between the two detectors, the values are in closer agreement using additional engineering parameters than for the systematic model outlined in Section~\ref{sec:eureka}, which does not consider engineering parameters. This, coupled with a reduction in the residual correlated noise when including engineering terms, suggests that further exploration is warranted of such engineering terms in future efforts to reduce correlated noise in similar observations using NIRSpec G395H.

Using these different reductions, we then assess the impact of the correlated noise (and timing offsets) on our spectrum. We next calculate and compare the transmission spectrum using our engineering systematic noise model and our nominal position and time noise models.

\subsection{Transmission Spectrum}
\begin{figure*}
\begin{centering}
\includegraphics[width=\textwidth]{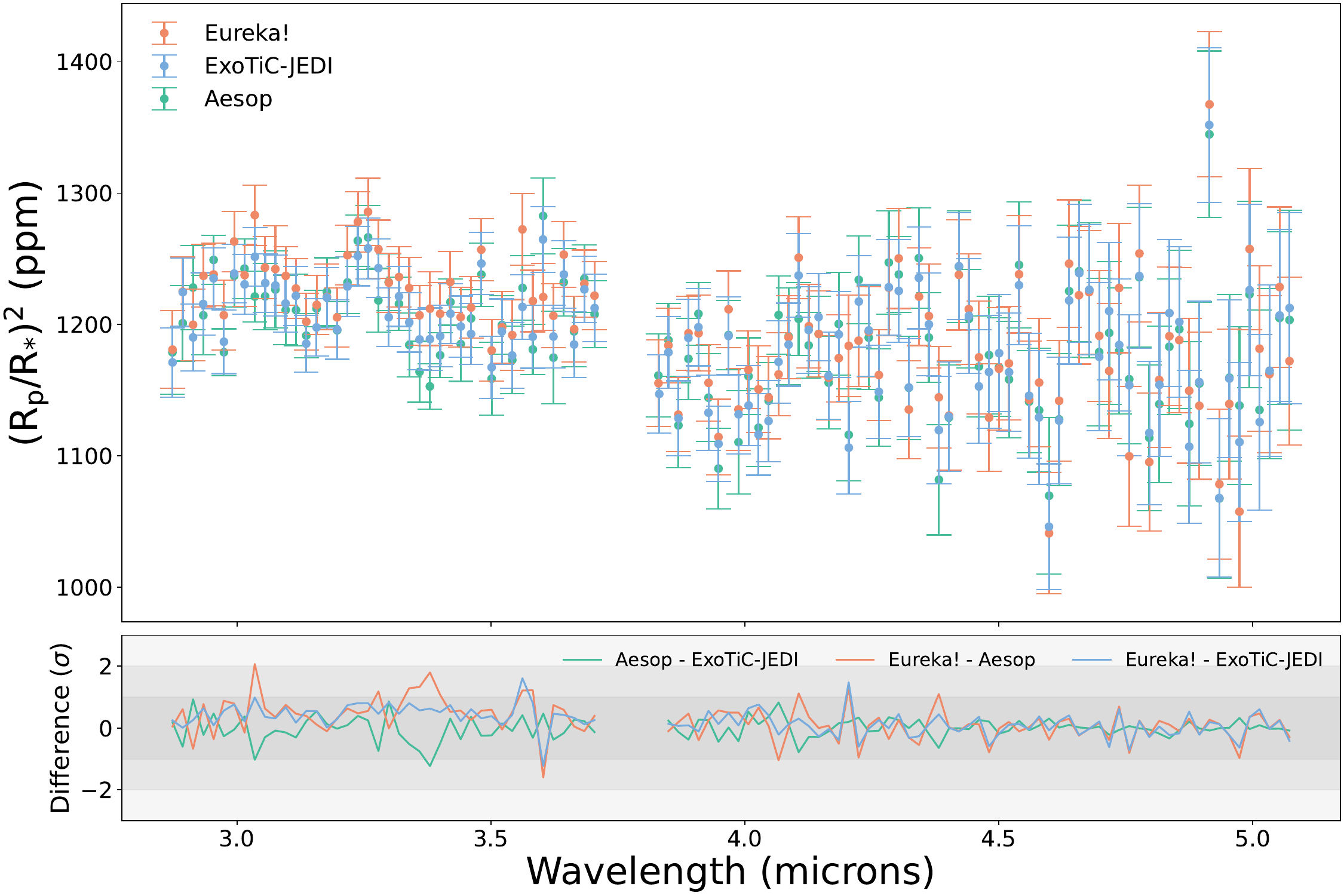}
\caption{We show the transmission spectrum for each reduction in the top panel and the differences between the reductions (in units of sigma) in the bottom panel. All three of our reductions agree per-point, at the $\sim$2$\sigma$ level or better, with a median difference of 6ppm.} 
\label{figure:spectrum}   
\end{centering}
\end{figure*}

Despite the residual correlated noise present in the fits to the white light curves, there did not seem to be a substantial amount of residual correlated noise in the spectroscopic bins (see Figure~\ref{figure:bins_rms}), so we first determine the transmission spectrum using each of our three reductions with the nominal (not including engineering mnemonics) noise models described in Section~\ref{sec:reductions}. We find that all three of our reductions agree per-point at the $\sim$2$\sigma$ level or better (Figure~\ref{figure:spectrum}). This indicates that individual reduction and fitting specific choices do not seem to greatly impact our derived transmission spectrum. All three of our reductions are consistent with a relatively flat spectrum with no easily identifiable features.

We also generate a transmission spectrum using our 13 parameter best-fitting engineering model (described in Section~\ref{sec:engineering}), which reduced the correlated noise in both NRS1 and NRS2. This exercise is important in determining the robustness of our spectrum and any constraints that we can place on the planet's atmosphere. 

\begin{figure*}
\begin{centering}
\includegraphics[height=.65\textwidth]{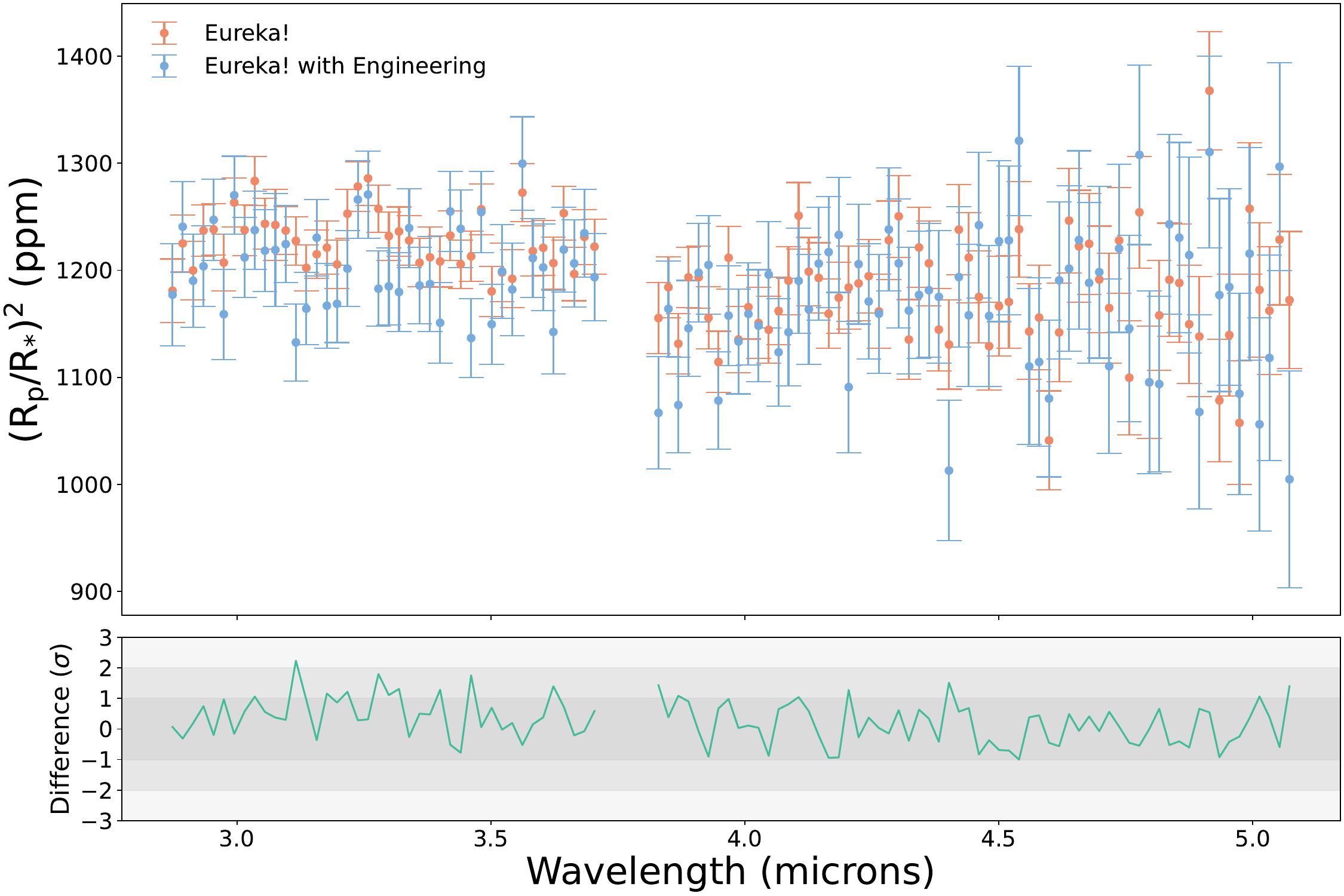}
\caption{We show the transmission spectrum for the nominal \texttt{Eureka!} case (detailed in Section~\ref{sec:eureka}) and the spectrum where we utilize a 13 engineering parameter systematic noise model. We show the differences between the nominal spectrum and spectrum that included engineering parameters (in units of sigma) in the bottom panel. The two spectra agree per-point at the 2.3$\sigma$ level or better with a median difference of 17ppm.}
\label{figure:spectrum_engineering}   
\end{centering}
\end{figure*}

We find that all of the bins in the transmission spectrum generated with and without the engineering parameters agree at the 2.3$\sigma$ level or better, indicating that the noise that is likely affecting the transit times does not alter our spectrum significantly (Figure~\ref{figure:spectrum_engineering}). This, along with the lack of residual correlated noise in the spectroscopically binned light curves (Figure~\ref{figure:bins_rms}), indicates that our transmission spectrum is robust despite the timing offset between the white light curves from the two detectors. Additionally, we extracted the spectrum for our \texttt{Eureka!} reduction using only the MCMC chains from each detector separately as priors (i.e. at the two different transit times) and find that the per-point transmission spectra using the best-fit NRS1 time and the best-fit NRS2 time each agree with the nominal spectrum at the 0.08$\sigma$ level or better. Since our nominal \texttt{Eureka!} fits utilize wide Gaussian priors derived from each chain (of 3$\times$ the standard deviation in the combined chains), this further shows that neither our choice of wide Gaussian prior nor the bimodal times alter the resulting transmission spectrum. Since all of our transmission spectra agree regardless of the inclusion of engineering parameters or prior on the time of transit, for the \texttt{Eureka!} reduction, we focus our subsequent analysis on the spectrum derived without engineering parameters, using the wide combined chain priors for each parameter when modeling the atmosphere. 

\section{Models}\label{sec:models}

In order to interpret our transmission spectrum, we employ three different techniques on all three data reductions. First, we fit for a series of ``non-physical models'' to understand the overall shape of the spectra and determine the significance of any structure that appears to be a feature. Second, we computed a grid of ``fixed-climate spectral models'' using \texttt{PICASO} \citep{batalha2019} with a fixed parameterized temperature-pressure profile for a range of metallicities and grey-cloud slab pressure levels. This gives us an overview of the atmospheric mean molecular weight and opaque pressure level allowed by our data. To give context to the ``non-physical models'' and the ``fixed-climate spectral models'', we also compute a small grid of radiative-convective (RC) models using \texttt{PICASO} 3.1 \citep{Mukherjee_2023} that allows us to gain a deeper understanding of the potential climate structure of TOI-836c. Finally, using these RC temperature-pressure profiles we leverage a microphysical model \citep[\texttt{CARMA};][]{Gao2023} to compute an altitude-dependent haze profile. Ultimately, this allows us to put into context the physical plausibility of an optically thick aerosol at low pressures that could be impeding the detection of strong molecular features. 

\subsection{Non-Physical Models}\label{sec:nonphys}

\begin{table*}
\caption{The results from our seven non-physical models for each of our three reductions. We show the log likelihood, the $\chi^{2}$/N, and the offset ($=$NRS1$-$NRS2) between the detectors (where applicable) for each model.}
\begin{tabular}{l|ccc|ccclll}
& \multicolumn{3}{c|}{\textbf{Eureka!}}& \multicolumn{3}{c|}{\textbf{Exo-TiC-JEDI}}& \multicolumn{3}{c}{\textbf{Aesop}} \\ \hline
\textbf{Model Type}& log Z& $\chi^2/N$   & Offset  {[}ppm{]}& log Z& $\chi^2/N$& \multicolumn{1}{c|}{Offset  {[}ppm{]}}& log Z   & $\chi^2/N$   & Offset{[}ppm{]}
\\ \hline
Slope Zero& -106 & 1.92 & N/A& -91  & 1.64 & \multicolumn{1}{c|}{N/A}  & -96& 1.74& N/A\\
Step Function  & -72  & 1.24 & 51& -70  & 1.19 & \multicolumn{1}{c|}{41}& -83& 1.43& 34 \\
Line   & -79  & 1.38 & N/A& -77  & 1.33 & \multicolumn{1}{c|}{N/A}  & -85& 1.48& N/A\\
Gaussian  & -78  & 1.25 & N/A& -77  & 1.23 & \multicolumn{1}{c|}{N/A}  & -87& 1.45& N/A\\
NRS1 Gaussian & -72  & 1.21 & 48& -71  & 1.18 & \multicolumn{1}{c|}{43}& -82& 1.32& 31 \\
NRS2 Gaussian & -72  & 1.23 & 53& -70  & 1.19 & \multicolumn{1}{c|}{43}& -83& 1.42& 35 \\
Gaussian + Line   & -78  & 1.31 & N/A& -77  & 1.30 & \multicolumn{1}{c|}{N/A}  & -85& 1.44& N/A  \\

\end{tabular}
\label{tab:nonphysical}
\end{table*}

\begin{figure*}
\begin{centering}
\includegraphics[width=.95\textwidth]{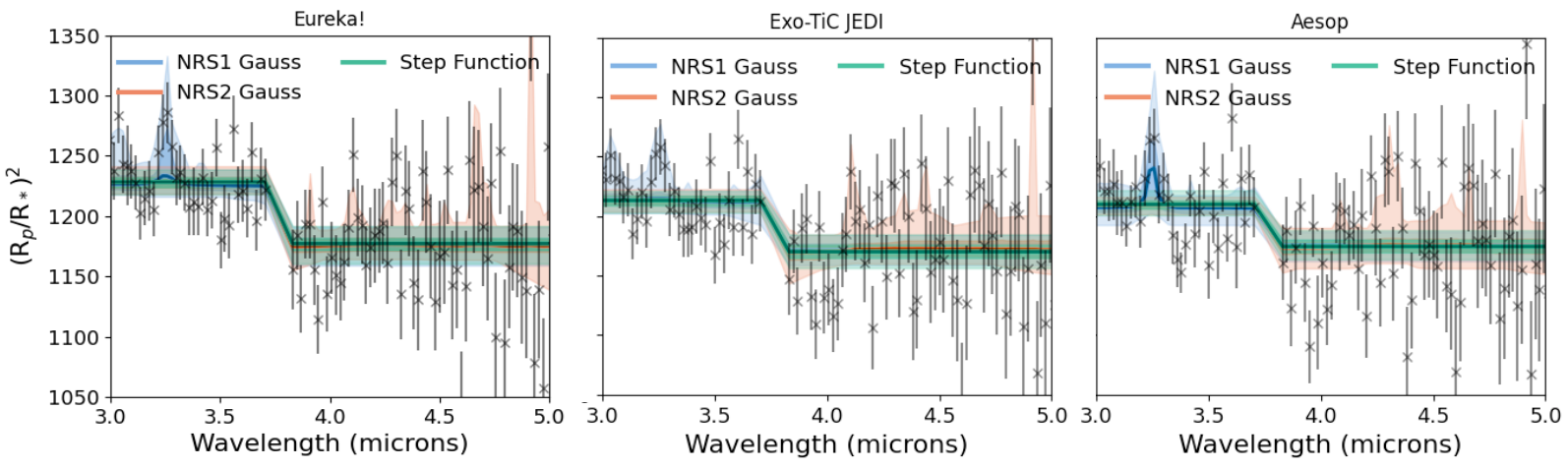}

\caption{We show the transmission spectrum for the nominal \texttt{Eureka!} reduction  (detailed in Section~\ref{sec:eureka}), the \texttt{ExoTiC-JEDI}, and \texttt{Aesop} reductions, respectively compared to the fitting results of the three highest-likelihood non-physical models. The three highest-likelihood non-physical models are consistent between reductions. In all panels, the solid colored line indicates the  median fit from the samples, the dark shaded region is the 68-percentile region (1$\sigma$) and the light shaded region is the 99-percentile region. The Bayes factor, as defined by the difference in evidence, for each of the non-physical models prefers the step function model in all three reductions. }
\label{figure:nonphysical}   
\end{centering}
\end{figure*} 

Using the ``MLFriends'' Nested Sampling Algorithm \citep{MLFriends2016, MLFriends2019} implemented in the open source UltraNest code \citep{Ultranest} we test seven different non-physical models, which contain increasing levels of complexity. The model suite is as follows: 

\begin{itemize}
    \item Model 1 [Zero-sloped line, 1 free parameter]: Fits for the y-intercept assuming zero slope. 
    \item Model 2 [Step function, 2 free parameters]: Fits for two y-intercepts independently for NRS1 and NRS2. Assumes zero slope for both.
    \item Model 3 [Line, 2 free parameters]: Fits for the slope and the y-intercept. 
    \item Model 4 [Gaussian, 4 free parameters]: Tests for the case of a single Gaussian-shaped spectral feature. Multiple features, if they were significant, are captured via degenerate solution. This fits for $y = c + A e^{(\lambda - \lambda_0)^2/\sigma^2}$, where $c$ is the baseline, $A$ is the amplitude of the feature, $\lambda_0$ is the wavelength midpoint of the feature, and $\sigma$ is the width of the feature.
    \item Model 5 [NRS1 Gaussian + offset, 5 free parameters]: A superposition of Model 2 and 4 and therefore accounts for an offset between NRS1 and NRS2 while allowing for a Gaussian feature in the NRS1 wavelength region. The prior on $\lambda_0$ is constrained to that of NRS1.
    \item Model 6 [NRS2 Gaussian + offset, 5 free parameters]: Accounts for an offset between NRS1 and NRS2 while allowing for a Gaussian feature in the NRS2 wavelength region.
     \item Model 7 [Gaussian + line, 5 free parameters]: A superposition of Model 3 and 4 and therefore a Gaussian feature with a sloped line. This would account for any changes to a continuum, along with a feature.  
\end{itemize}

Given the resultant evidence of seven model runs, we compute the Bayes factor ($\ln$B$_{12}=\ln$Z$_1$\,[Model 1] $-$ $\ln$Z$_2$\,[Model 2]; \cite{Trotta2008}) in order to determine if one model is preferred over another. The Bayes factor only directly translates to a $\sigma$-significance (e.g., Eqn. 27 in \citet{Trotta2008}) if the model's free parameters are nested (e.g. $f(c_1,c_2,c_3=A,...,c_N=Z)$ vs. $f(c_1,c_2,c_3,...,c_N)$, where A and Z are constants less than N). Therefore in order to compare all models, we focus on the Bayes factor. Figure \ref{figure:nonphysical} shows the results of the three highest likelihood models (step function, and NRS1 or NRS2 Gaussian + offset) for \texttt{Eureka!}, \texttt{ExoTiC-JEDI}, and \texttt{Aesop} reductions, respectively, from left to right. The full numerical results are shown in Table \ref{tab:nonphysical}. Overall, the fitting results are consistent between reductions within 1$\sigma$ for each of the fit parameters. For all reductions, the step function and either the NRS1 or NRS2 Gaussian + offset models are all strongly preferred ($|\ln$B$_{12}|>13$) when compared to the slope-zero line. However, when comparing the simple step function model against any of the Gaussian models, none of the latter are preferred for any of the reductions because the increase in number of free parameters does not improve the evidence despite minor improvements to the $\chi^2$/N. This suggests that the transmission spectrum, regardless of data reduction method, is well-explained by two zero-sloped lines (step function) with an offset of 34, 41, and 51~ppm for the \texttt{Aesop}, \texttt{ExoTiC-JEDI}, and \texttt{Eureka!} reductions, respectively. The tentative bump at 3.25$\mu$m appears within 1$\sigma$ of the highest likelihood solution of NRS1 Gaussian + offset model and the Gaussian + line model for both reductions. However, when compared with a featureless offset model, it is not preferred (ratio of the model evidences is $<1$). Other even more tentative features in NRS1 and NRS2 shown within the 99-percentile region of Figure \ref{figure:nonphysical} are likely due to systematic scatter. Similarly, when compared with the step function model, these features are not preferred because the increase in likelihood is not enough to outweigh the increase in free parameters. Even though the scatter seems to be consistent between reductions, additional visits of TOI-836c would be needed to determine the origin of this scatter. 

\subsection{Fixed-Climate Spectral Models}

\begin{figure*}
\begin{centering}
\gridline{\fig{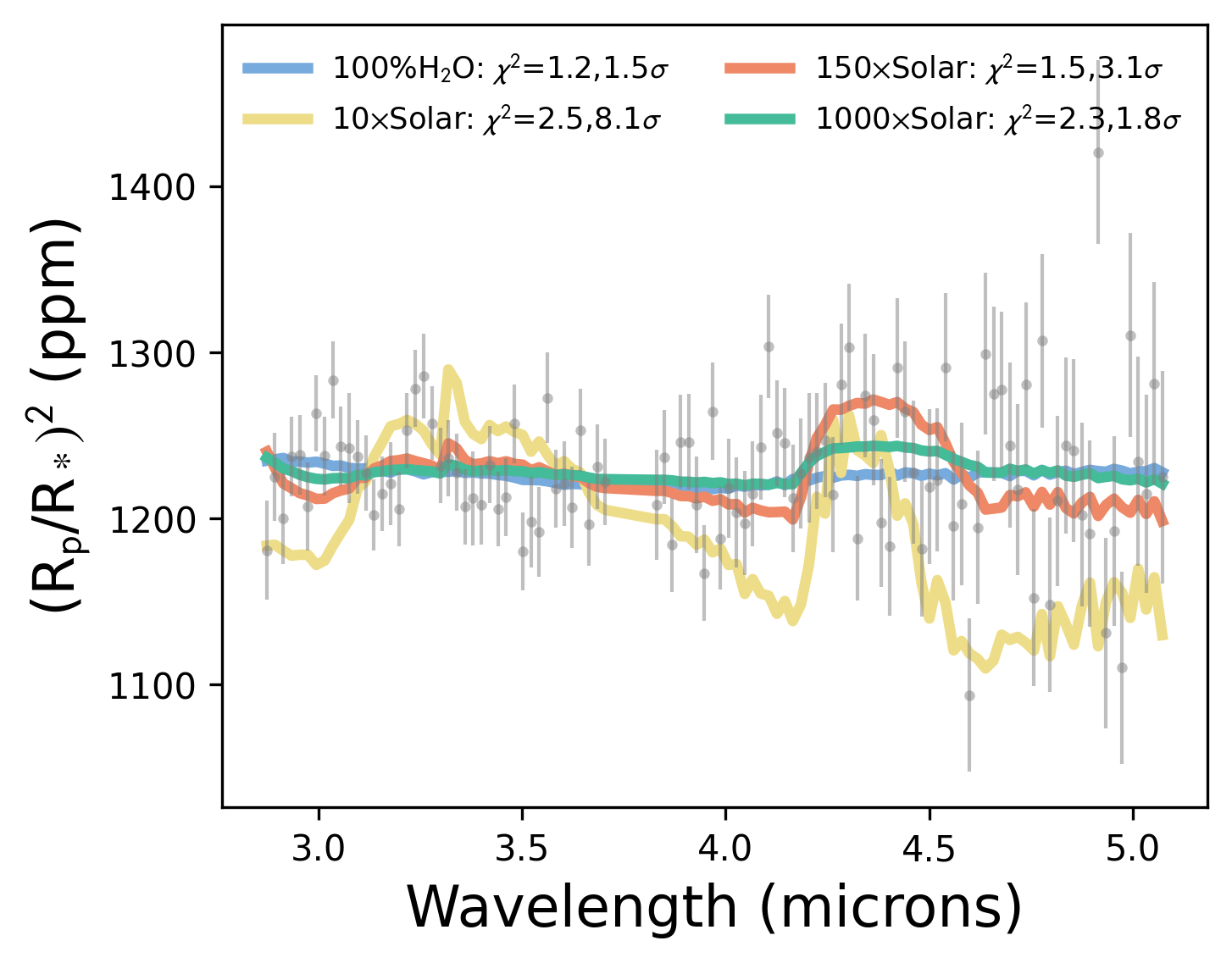}{0.45\textwidth}{(a)}
\fig{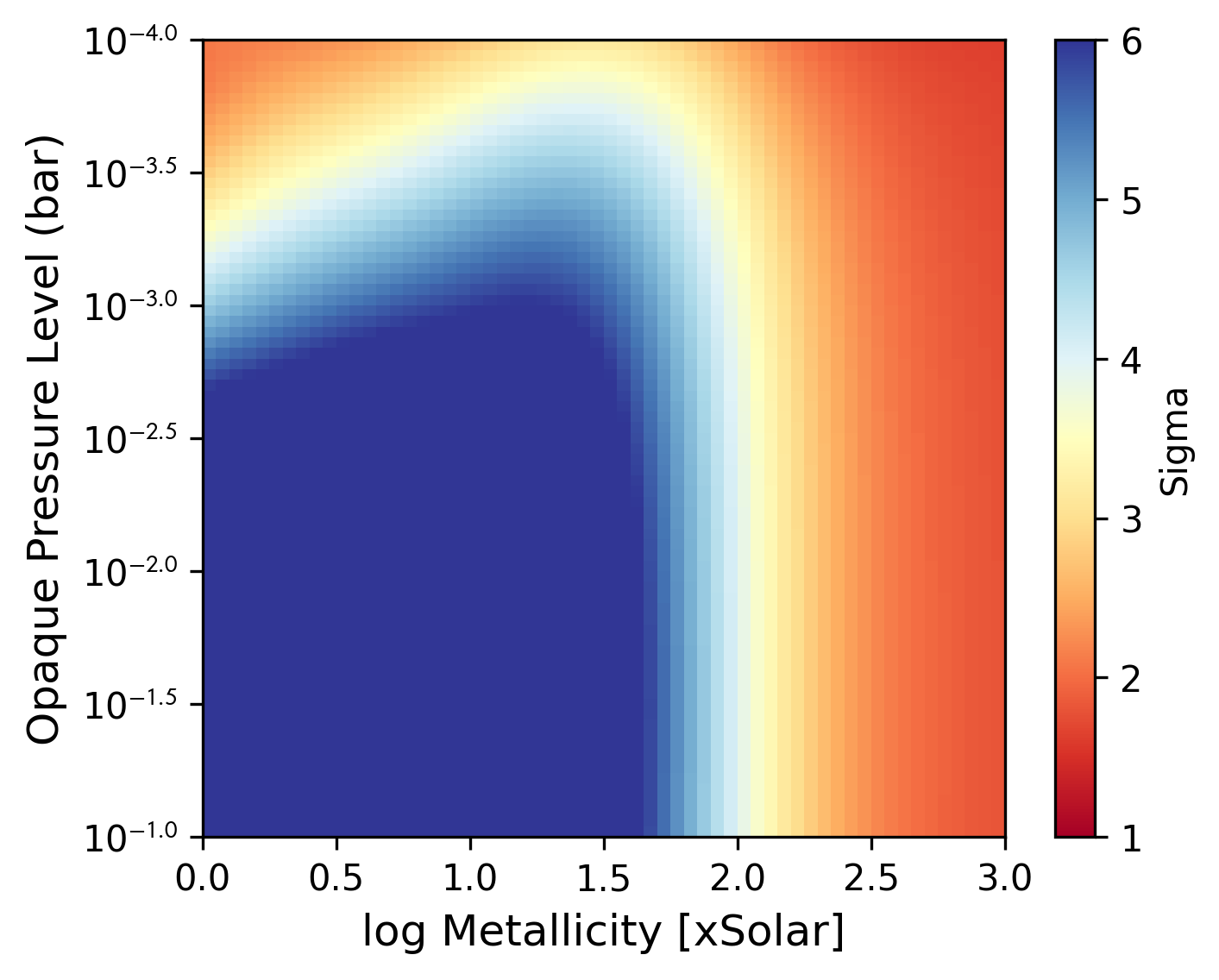}{0.45\textwidth}{(b)}}
\caption{We show the transmission spectrum for the nominal \texttt{Eureka!} case (as detailed in Section~\ref{sec:eureka}) along with four example models depicting 10$\times$, 150$\times$, and 1000$\times$ Solar metallicity, as well as a 100\% H$_2$O model. In the legend, we also show the $\chi^2/N$ and the corresponding $\sigma$ at which the model can be confidently ruled out. On the left (b) we show the full parameter space of opaque pressure level and metallicity. Here, the color scale of the heatmap similarly indicates at what $\sigma$ each model can be confidently-ruled out. Ultimately, we are unable to rule out metallicities $>$175$\times$ Solar (log[M/H]$>$2.2).}
\label{fig:phys}   
\end{centering}
\end{figure*} 

The non-physical model results showcase the agreement between all three data reduction methods. Therefore, in this section, we show spectral models compared against the nominal \texttt{Eureka!} reduction, without the addition of the engineering parameters. The ultimate goal here is to determine what range of mean molecular weights and ``opaque pressure levels'' can be ruled out. Here, we use metallicity as a proxy for tuning the mean molecular weight of the atmosphere. The ``opaque pressure level'' is a term adapted from the term ``apparent surface pressure'' that was used in \citetalias{LustigYaeger2023} and term ``cloud top pressure'' that is frequently used in previous studies of exoplanets \citep[e.g.][]{Kreidberg2014, Knutson2014}. All these terms generally define a grey opacity below which our measurements are not able to probe. The simplest way to implement this is to imagine a grey optically thick cloud deck.  

For the fixed-temperature-pressure profile, we use a 1D, 5-parameter double-grey analytic formula \citep{guillot2010}, which for weakly irradiated systems approximates to $T^4 \sim 0.75 * T_\mathrm{eqt}^4 \left( p \mathrm{[bar]} + 2/3)\right)$. This is similarly adopted in several other works that aim to approximate the climate structure when in-depth modeling is not warranted by the precision of the data \citepalias[e.g.,][]{LustigYaeger2023, Moran2023, May2023}. Given the climate structure, we compute atmospheric abundance profiles assuming chemical equilibrium using the chemical equilibrium grid computed by \citet{Line2013} with NASA's CEA code \citep{gordon1994computer}. The chemistry grid is publicly available on \texttt{GitHub} as part of \texttt{CHIMERA}'s open source code\footnote{https://github.com/mrline/CHIMERA}. The grid computes the abundances of 19 molecules (H$_2$O, CH$_4$, CO, CO$_2$, NH$_3$, N$_2$, HCN, H$_2$S, PH$_3$, C$_2$H$_2$, C$_2$H$_6$, Na, K, TiO, VO, F, H, H2, He) at 26 metallicities (log [M/H]=-2--3), and 25 C/O ratios (C/O = 0.01--2). We choose a fixed solar C/O ratio (0.55) in accordance with \citet{Asplund2009} because we are interested in leveraging metallicity as a proxy for changes in mean molecular weight. While changes in C/O can change the expected dominant molecular spectral features, we have demonstrated in \S\ref{sec:nonphys} that there are no statistically significant features in any of the data reductions and thus an exploration of C/O variations is not warranted. Here we are simply interested in determining what range of mean molecular weights can be statistically ruled out.  Lastly, we implement the cloud deck as an optically thick slab below a certain pressure (varied between 1--10$^{-4}$~bars). Given these three components (temperature-pressure profile, chemistry, cloud) we use the \texttt{PICASO} radiative transfer code \citep{batalha2019} to compute transmission spectra between 3--5~$\mu$m to compare to our NIRSpec G395H data. We rebin our high resolution spectral models (R$=$60,000) to the wavelength grid shown in Figure \ref{figure:spectrum}. 

Figure \ref{fig:phys} shows the result of this analysis, which shows we can generally rule out  atmospheric compositions with [M/H]$<175\times$Solar (mean molecular weight$\sim$6). In Figure \ref{fig:phys}a we show three chemical equilibrium models (10$\times$, 150$\times$, and 1000$\times$), all of which have a grey opacity at 0.1 bar.  We additionally show a 100\% H$_2$O model, which cannot be ruled out, for reference. In this analysis we fit for detector offsets individually to each model, before computing the $\chi^2/N$. However, in the figure we add these offsets directly to the data in order to better visualize the model. For a visual representation of the offsets, see Figure \ref{figure:nonphysical}.  Shown in Figure \ref{fig:phys}b is the full parameter space explored. Above a metallicity of $175\times$ Solar (mean molecular wight$\sim$6) we lose the ability to rule out physical cases at 3$\sigma$. This result is relatively insensitive to the optical depth of the cloud slab until pressures $<$10$^{-4}$ bar, where we are unable to rule out even Solar metallicity cases. In the following section, we conduct more in-depth modeling to understand whether it is physically plausible to form aerosols in the atmosphere of TOI-836c at such low pressures. 

\subsection{Haze Modeling with CARMA}

Our ability to rule out metallicity cases less than $175\times$ is dependent on the location of an optically thick aerosol layer. Therefore, here we leverage a microphysical model (CARMA) to determine if an optically thick aerosol layer is physically plausible below 10$^{-4}$ bar.   First, we generated self-consistent climate models in order to more closely investigate the physics of the planet's atmosphere using \texttt{PICASO} 3.1 \citep{batalha2019, Mukherjee_2023}. By ``self-consistent,'' we mean that the model is initiated with a parametric pressure-temperature profile and chemical equilibrium abundances, then iterated until it achieves radiative-convective equilibrium. In our models, we assume full day-to-night heat redistribution. 
For this use case, \texttt{PICASO} supports climate calculations up to  100$\times$ Solar metallicity. We select the 100$\times$ Solar case as a representative case to probe the physical plausibility for high-altitude aerosols. 

We use CARMA \citep[Community Aerosol and Radiation Model for Atmospheres;][]{turco1979,toon1988,jacobson1994,ackerman1995} to simulate the optical depth of photochemical hazes in the atmosphere of TOI-836c, with the 100$\times$ Solar metallicity model atmosphere described in the previous paragraph as a nominal background. CARMA is a 1D aerosol microphysics code that accounts for the nucleation, condensation, coagulation, evaporation, sedimentation, diffusion, and advection of cloud and haze particles in planetary atmospheres \citep{gao2018carma}. Here we apply it in a similar way as in \citet{Gao2023}, where we compute the vertical and size distribution of photochemical hazes given a range of column haze production rates (10$^{-14}$--10$^{-9}$ g cm$^{-2}$ s$^{-1}$ as informed by sub-Neptune photochemical models, e.g. \citealt{kawashima2019,lavvas2019}). Briefly, we assume the production of 10 nm spherical ``seed'' particles with internal density of 1 g cm$^{-3}$ at pressure levels $<$1 $\mu$bar, which then settle downwards in the atmosphere under the effect of gravity and vertical mixing. We parameterize vertical mixing using eddy diffusion, with the eddy diffusion coefficient derived for a 100$\times$ Solar metallicity atmosphere from \citet{charnay2015kzz}. As the particles sediment, they are allowed to grow through collisions (coagulation), with the assumption that they remain spherical. 

\begin{figure}
\begin{centering}
\includegraphics[width=.49\textwidth]{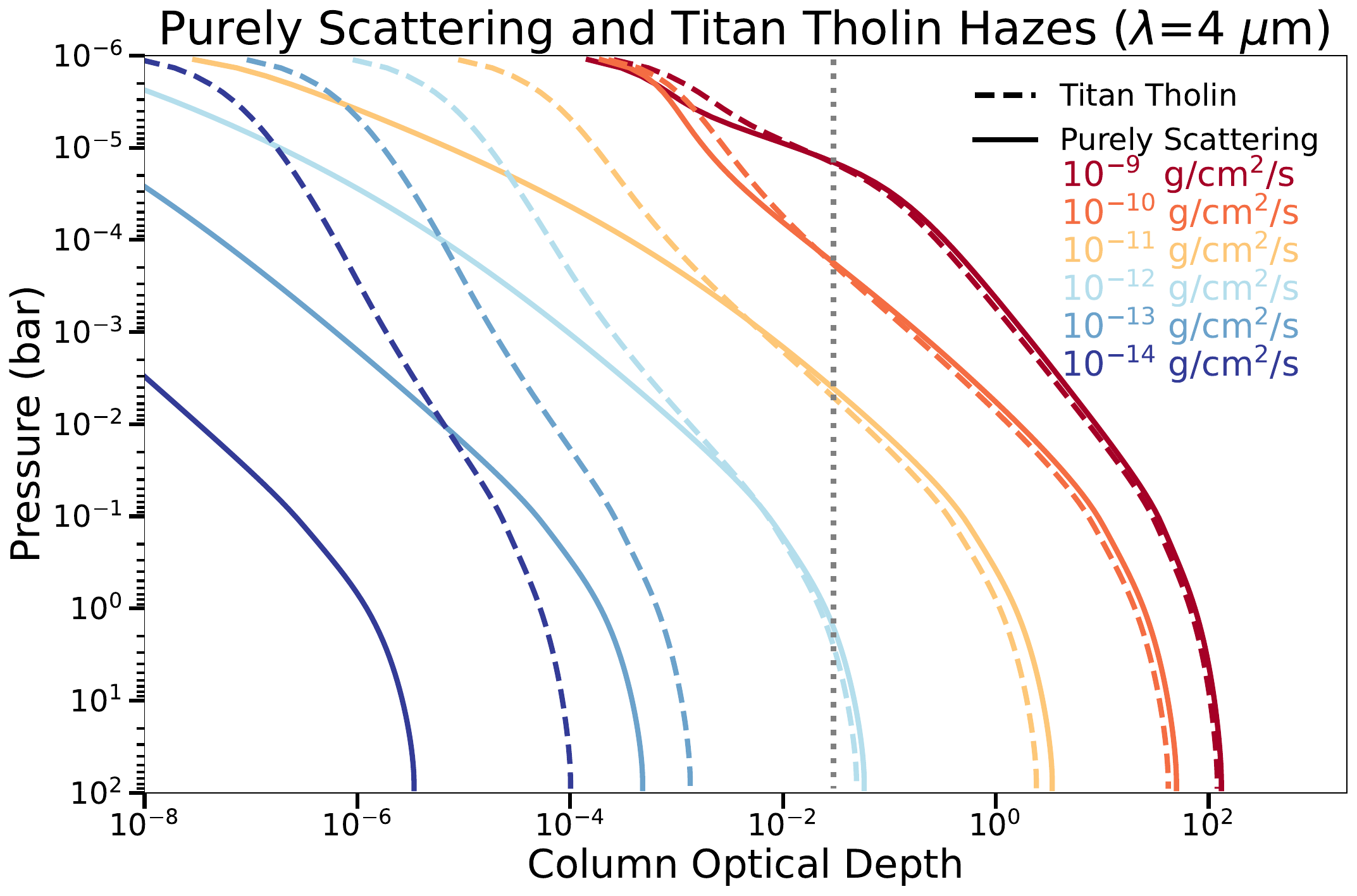}
\caption{Results from our CARMA modeling of TOI-836c. The curves show the cumulative nadir optical depth at a wavelength of 4 $\mu$m for different haze production rates and assuming purely reflective haze optical properties (solid lines) and tholin haze optical properties (dashed lines). The vertical dotted line shows which nadir optical depth gives a slant optical depth of 1 for a 100$\times$ Solar metallicity atmosphere following \citealt{fortney2005}, with a mean molecular weight of 4.25 g/mol.}
\label{fig:haze}   
\end{centering}
\end{figure}

We compute the optical depth per layer of the resulting model haze distributions assuming Mie scattering and both the purely scattering and Titan tholin refractive indices \citep{Khare1984} as in \citet{Gao2023}. We then sum up the layer optical depths to arrive at the nadir cumulative optical depth, shown in Figure \ref{fig:haze} for the different haze production rates and haze refractive indices. To compute the slant optical depth, we use Eq. 6 from \citet{fortney2005} and assume a mean molecular weight of 4.25 g mol$^{-1}$ for the 100$\times$ Solar metallicity atmosphere. We find that column haze production rates $>$10$^{-10}$ g cm$^{-2}$ s$^{-1}$ are needed to produce haze slant optical depth unity pressure levels $<$10$^{-4}$ bar. These haze production rates are at the upper limits of those predicted by photochemical models that source haze from the photolysis of atmospheric hydrocarbon and nitrogen species \citep{kawashima2019,lavvas2019}, but are not implausible, and are on par with those predicted for GJ 1214b based on its near-to-mid-infrared transmission spectrum \citep{Gao2023}. As such, TOI-836c may have an atmosphere with similar chemistry to GJ 1214b that is conducive to haze formation and an extremely high metallicity of $\geq$1000$\times$ Solar \citep{Kempton2023}.

\section{Discussion}\label{sec:discussion}

\subsection{Contextualizing TOI-836c}
The only two sub-Neptune exoplanets with published JWST transmission spectra to date are GJ 1214b (8.17 M$_{\earth}$, 2.628 R$_{\earth}$, 553K;  \citealt{Gao2023}) and K2-18b (8.63 M$_{\earth}$, 2.61 R$_{\earth}$, $\sim$250-300K; \citealt{Madhu2023}). TOI-836c (9.6 M$_{\earth}$, 2.587 R$_{\earth}$, 665K; \citealt{Hawthorn2022}) is more massive and hotter than both of these planets, but has a smaller radius, meaning that it has the highest gravity. Our observations of TOI-836c rule out atmospheric metallicities $<$175$\times$ Solar in the absence of clouds or hazes, which is similar to the high metallicities inferred for both GJ 1214b and K2-18b. Transmission spectra of GJ 1214b from near- and mid-infrared observations (including the aforementioned JWST MIRI LRS observation), are consistent with a metallicity of $>$300$\times$ Solar and prefer a metallicity of $\geq$ 1000$\times$ Solar with a haze production rate of $\geq$ 10$^{-10}$ g cm$^{-2}$ s$^{-1}$ \citep{Gao2023}. Such a haze production rate is also consistent with our observations of TOI-836c, as we are unable to rule out a cloudy or hazy atmosphere. Transmission spectra of K2-18b using both JWST NIRISS SOSS and NIRSpec G395H are consistent with a $\sim$100$\times$ Solar composition \citep{Wogan2024}, a value  smaller than our inferred lower bound for TOI-836c with a clear atmosphere. Using these same spectra, \cite{Madhu2023} interpreted the composition of K2-18b's atmosphere to be indicative of the presence of a deep sub-surface ocean of liquid water, following published photochemical simulations \citep[e.g.,][]{Hu2021}. If that is the case, volatile elements seen in the atmosphere are only a fraction of the total budget of volatiles in the planetary bulk.

If TOI-836c does have a high metallicity atmosphere and K2-18b is consistent with a 100$\times$ Solar sub-Neptune, then all three sub-Neptunes observations published thus far with JWST have relatively high atmospheric metallicities and would be comparable to the ice giants of our own Solar System. The most recent estimates of the methane abundance in the atmospheres of Uranus and Neptune indicate deep methane volume mixing ratios of 2.3--4.0\% (C/H 47--80$\times$ Solar) in Uranus, and 2.2--2.9\% (C/H 45--60$\times$ Solar) in Neptune, derived from the methane abundance below the methane cloud at $\sim 1.1$ bar \citep[][respectively]{Sromovsky2019,Tollefson2019}. On the other hand, the total volatile content of Uranus and Neptune's envelope, i.e. excluding the rocky core, inferred by interior structure models is 75\% to 85\% water by mass \citep{Nettelmann2013,Podolak2019}. These values correspond to average metallicities of the envelope of 200--380$\times$ Solar. The existence of a two-layer envelope with two distinct metallicities is necessary to explain the gravitational data from Voyager 2 flybys of Uranus and Neptune, as well as ring seismology measurements \citep{Elliot1984,Jacobson2006,Jacobson2009}. This structure is usually attributed to the rain-out of water, forming a deep layer of super-ionic water with a hydrogen-rich outer envelope on top. It has been suggested that terrestrial planets that receive insolations of more than $\sim 1.06$ times the present-day Earth value are expected to be in a post-runaway greenhouse stage, leading to the formation of steam atmospheres that are fully-mixed \citep{Kopparapu2013,Turbet2019}. Therefore, a hot and fully-mixed analog of Uranus or Neptune would have a volatile content comparable to our $175\times$ Solar metallicity constraint if TOI-836c has a clear atmosphere. However, it is unclear whether the post-runaway greenhouse effect would lead to fully-mixed envelopes in sub-Neptunes \citep{Pierrehumbert2023}.

It is possible that TOI-836c, and most sub-Neptunes, share additional similarities with the Solar System ice giants from a formation perspective. \cite{Lambrechts2014} proposed that the difference in bulk compositions between ice giants and gas giants is due to a premature halting of the growth of ice giants. According to this scenario, after the cores of giant planets were formed, the embryos accreted a mixture of 20\% nebular gas and 80\% of ice-rich pebbles by pebble accretion \citep{Lambrechts2012}. After the reservoir of pebbles was exhausted, gas giants continued to accrete nebular gas, diluting their volatile rich envelopes down to the low metallicities observed today \citep[$\sim 3$ and $\sim 10$$\times$ Solar for Jupiter and Saturn, respectively;][]{Mousis2018,Atreya2020}. Ice giants reached the pebble isolation mass and stopped their growth, resulting in a final composition of $\sim 20\%$ H$_2$-He gas, and $\sim 80\%$ heavy elements for their envelopes. Such compositions can also be achieved by accretion of planetesimals \citep[e.g.][]{Fortney2013}, where the composition of the envelopes depends on the size of accreted planetesimals. Determining whether sub-Neptunes are smaller analogs of ice giants or if they have distinct compositions is key to improving our understanding of planetary formation, and highlights the importance of measuring fundamental properties such as atmospheric metallicities.

While our observations of TOI-836c seem to be consistent with the high metallicities inferred for the other sub-Neptunes in the absence of high altitude clouds, both K2-18b and GJ 1214b show direct evidence for molecular features. Both the dayside and nightside spectra of GJ 1214b are inconsistent with blackbody expectations at $>$3$\sigma$ \citep{Kempton2023}, while K2-18b shows direct evidence of CH$_{4}$ and CO$_{2}$. However, TOI-836c is both hotter and has a higher gravity than these other two sub-Neptunes, possibly explaining the lack of detected features at current S/N in the transmission spectrum of this planet. To further contextualize this planet, we next model the bulk composition of TOI-836c to determine its likelihood as a water world as has been previously argued for K2-18b.

\subsection{The Bulk Composition of TOI-836c}\label{sec:bulk}

As our NIRSpec observations are unable to place non-degenerate constraints on the atmosphere of TOI-836c, we instead place model-dependent limits on its structure. To infer the bulk composition of TOI-836c, we use the open source code \texttt{smint}\footnote{https://github.com/cpiaulet/smint} developed by \cite{Piaulet2021} to perform a MCMC retrieval of planetary bulk compositions from pre-computed grids of theoretical interior structure models. We consider two possible compositions for the interior: 1) an Earth-like core with a H\textsubscript{2}-He envelope of Solar metallicity \citep{Lopez2014}, and 2) a refractory core with a variable core mass fraction and a pure H\textsubscript{2}O envelope and atmosphere on top \citep{Aguichine2021}. These compositions represent end-member cases between an envelope that would form with a Solar-like composition, and a H\textsubscript{2}-He free envelope. While our observations do rule out a clear Solar metallicity atmosphere, a Solar metallicity end-member case is still possible given sufficient aerosol opacity. Our goal is to determine the range of possible bulk volatile contents in TOI-836c in the absence of such observational constraints. 

We find that TOI-836c can have an envelope mass fraction of $1.74^{+0.55}_{-0.48} \%$ in the case of Solar metallicity gas, or a water mass fraction of $52^{+15}_{-14} \%$ in the pure H\textsubscript{2}O case (see Figure~\ref{fig:interior}). Detailed results and methods to produce Figure~\ref{fig:interior} are provided in Appendix \ref{sec:appendix-interior}. A 50\% water and 50\% rocky composition is interesting as it represents the bulk composition of the solid building blocks of planets (dust, pebbles, and planetesimals) when all the icy material has condensed, i.e. in the coldest region of the protoplanetary disk \citep{zeng2019}. Despite the fact that most interior structure models use water as a proxy for all volatiles, the 50\% water and 50\% rock models match the masses and radii of icy moons of the Solar System \citep[e.g.][]{Sotin2007} and also point towards a potential population of water worlds orbiting M dwarfs (see \cite{luque2022}, although \cite{Rogers2023} provide an alternative explanation). In reality, all intermediate compositions are also possible. If the atmosphere and the envelope have a similar metallicity, then an atmospheric metallicity of $175\times$ Solar could corresponds to an interior that is 1.6 wt\% H$_2$-He, 4.1 wt\% H$_2$O, and 94.3 wt\% of an Earth-like core. Atmosphere molecular detections could narrow the range of possible bulk composition even further with information on elemental ratios such as C/O or C/N.

\begin{figure*}
\begin{centering}
\includegraphics[width=\textwidth]{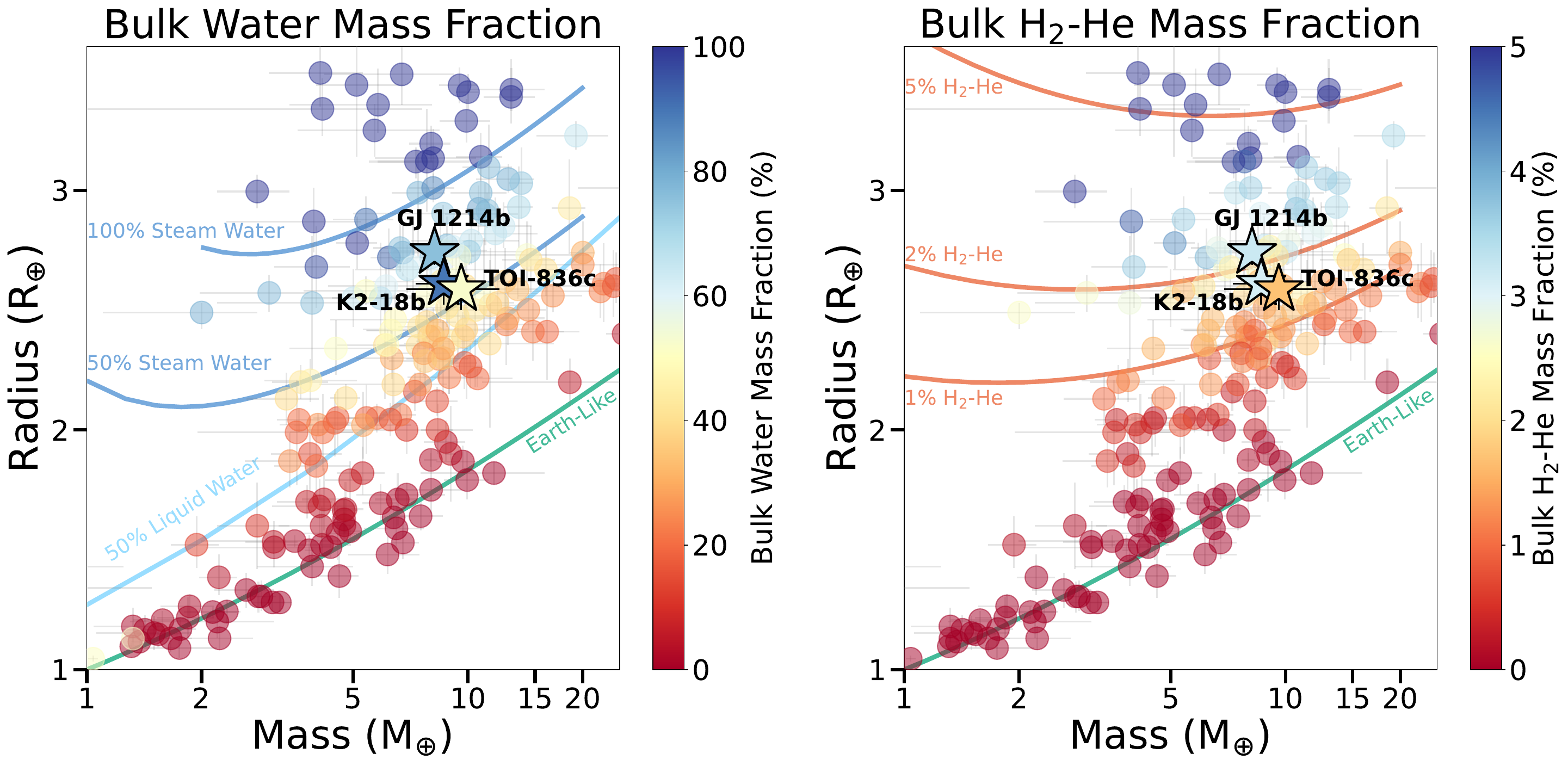}
\caption{Mass-radius diagram of small planets colored by bulk water mass fraction (left) and bulk H$_{2}$-He mass fraction (right). We show the sub-Neptunes with JWST observations (TOI-836c, GJ 1214b, and K2-18b) as stars and planets from the NASA Exoplanet Archive with mass errors of $<$20\% and radius errors of $<$5\% as circles. 
\cite{Zeng2016} curves are used for the Earth-like and 50\% liquid water compositions. \cite{Aguichine2021} curves are used for the 50\% and 100\% steam water composition, and are shown for $T_{\mathrm{eq}}=600$ K and an Earth-like core. \cite{Lopez2014} curves are used for the 1\%, 2\% and 5\% H$_2$-He composition, and are shown for $T_{\mathrm{eq}}=600$ K and an age of 5 Gyr. For each planet, the bulk volatile content (H$_2$O or H$_2$-He) has been determined with the method described in Appendix \ref{sec:appendix-interior}.}
\label{fig:interior}   
\end{centering}
\end{figure*} 

\section{Conclusions}\label{sec:conclusion}
We present the first JWST NIRSpec G395H transit observations of the sub-Neptune TOI-836c, the first planet observed as part of the COMPASS program. Using three independent reductions, we find a consistently featureless transmission spectrum. Our spectrum is able to rule out a clear low metallicity atmosphere ($<$ 175$\times$ Solar), as well as low metallicity atmospheres with clouds and hazes at pressures greater than $\sim$0.1-1 mbar. However, the planet likely has a sizable H\textsubscript{2}/He envelope as inferred from its mass and radius, and we show that the presence of clouds and hazes at low pressures cannot be ruled out, as in the case of GJ 1214b, which has a similar equilibrium temperature and mass. 

While our median measured transit depth precisions are only $\sim$1.13-1.18$\times$ those expected from \texttt{PandExo} for our three reductions, we do find a significant amount of residual correlated noise in the white light curves. We also find that the mid-transit times derived from the two different NIRSpec G395H detectors are inconsistent at $>$4$\sigma$ for this single simultaneous transit observation. We include mnemonics from the JWST Engineering Database in our systematic noise model and find that their inclusion can reduce both the residual correlated noise and the timing discrepancy to within 2.3$\sigma$, but does not fully remedy the offset or entirely remove the residual correlated noise. Nonetheless, further investigation of the use of mnemonics from the JWST Engineering Database may help to reduce residual correlated noise in NIRSpec G395H observations of small planets around bright stars.

While the results presented herein are only for the first planet observed as part of the COMPASS program, these results will be included in the entire COMPASS sample of 12 super-Earths and sub-Neptunes, which will be presented in forthcoming publications. Once the survey is complete, we will have a better understanding of the observability and compositions of the atmospheres of the most common type of planet in the Galaxy and provide crucial context for understanding the composition of planets in our own Solar System.

\vspace{\baselineskip}

This work is based on observations made with the NASA/ESA/CSA James Webb Space Telescope. The data were obtained from the Mikulski Archive for Space Telescopes at the Space Telescope Science Institute, which is operated by the Association of Universities for Research in Astronomy, Inc., under NASA contract NAS 5-03127 for JWST. These observations are associated with program \#2512. Support for program \#2512 was provided by NASA through a grant from the Space Telescope Science Institute, which is operated by the Association of Universities for Research in Astronomy, Inc., under NASA contract NAS 5-03127. This work is funded in part by the Alfred P. Sloan Foundation under grant G202114194. Support for this work was provided by NASA through grant 80NSSC19K0290 to JT and NW. This work benefited from the 2022 and 2023 Exoplanet Summer Program in the Other Worlds Laboratory (OWL) at the University of California, Santa Cruz, a program funded by the Heising-Simons Foundation. This material is based upon work supported by NASA’S Interdisciplinary Consortia for Astrobiology Research (NNH19ZDA001N-ICAR) under award number 19-ICAR19\_2-0041. We acknowledge use of the lux supercomputer at UC Santa Cruz, funded by NSF MRI grant AST 1828315. This research has made use of the NASA Exoplanet Archive, which is operated by the California Institute of Technology, under contract with the National Aeronautics and Space Administration under the Exoplanet Exploration Program. This paper makes use of data from the first public release of the WASP data \citep{Butters2010} as provided by the WASP consortium and services at the NASA Exoplanet Archive, which is operated by the California Institute of Technology, under contract with the National Aeronautics and Space Administration under the Exoplanet Exploration Program.

Co-Author contributions are as follows: 
NLW led the data analysis of this study. NEB led the atmospheric modeling efforts with contributions from NS and PG. LA, MKA, and JIAR  provided additional reductions and analyses of the data. AA provided interior models. All authors provided detailed comments and conversations that greatly improved the quality of the manuscript. 

\software{CARMA \citep{Gao2023}, \texttt{emcee} \citep{Foreman-Mackey2013}, \texttt{Eureka!} \citep{Bell2022},  \texttt{ExoTiC-Jedi} \citep{Alderson2022}, \texttt{ExoTiC-LD} \citep{Grant2022}, \texttt{Matplotlib} \citep{Hunter2007},  \texttt{NumPy} \citep{harris2020}, \texttt{PandExo} \citep{Batalha2017}, \citep{Virtanen2020}, \texttt{PICASO} \citep{batalha2019},  \texttt{SMINT} \citep{Piaulet2021}, \texttt{Tiberius} \citep{{Kirk2017, Kirk2021}}, \texttt{ultranest}\citep{Ultranest}}

\bibliography{main}

\begin{appendix}
\setcounter{figure}{0}
\renewcommand\thefigure{A\arabic{figure}} 

\section{Results from interior structure modeling} \label{sec:appendix-interior}

\begin{figure*}[h!]
\begin{centering}

\includegraphics[width=0.37\textwidth]{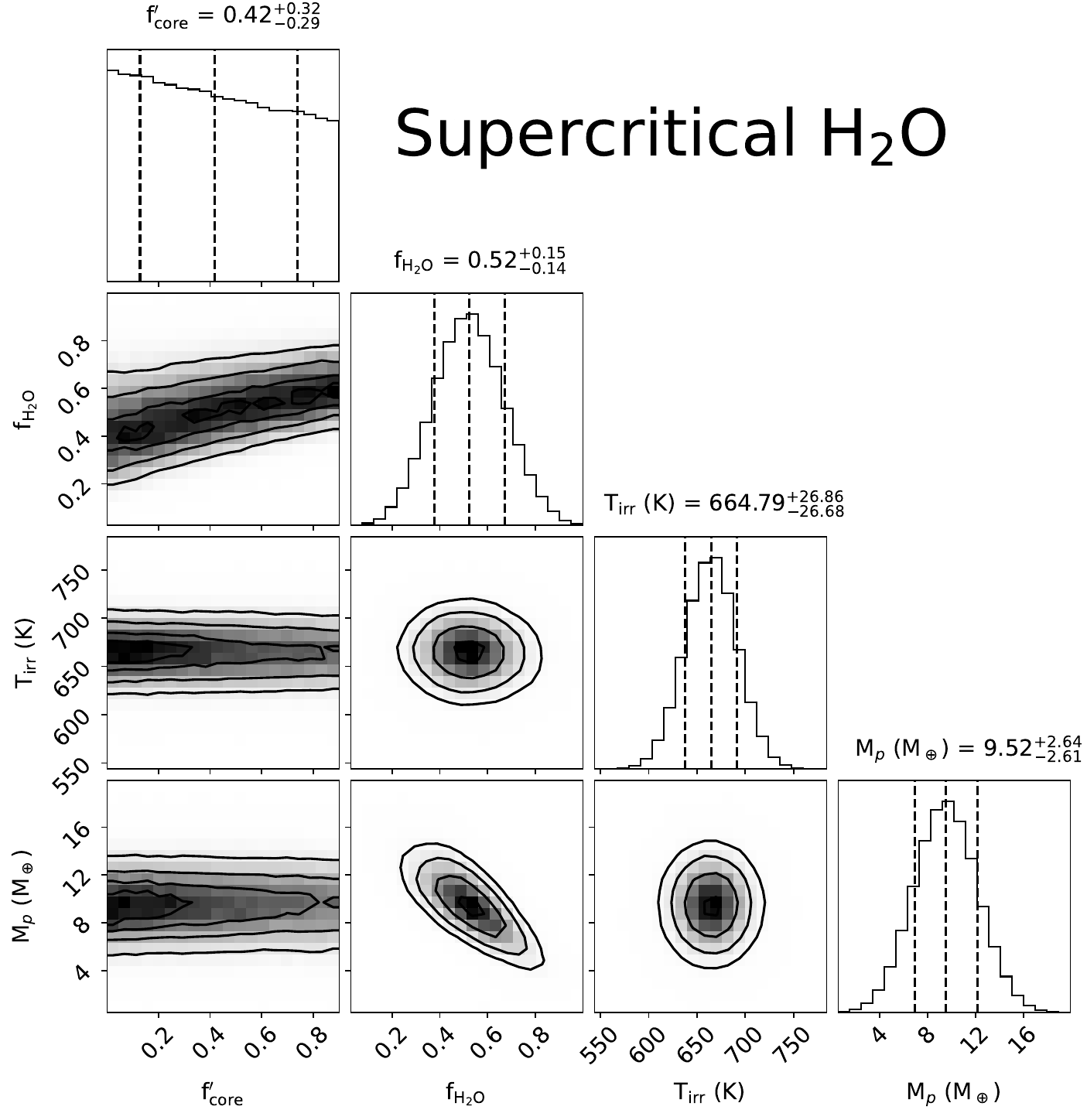} \includegraphics[width=0.37\textwidth]{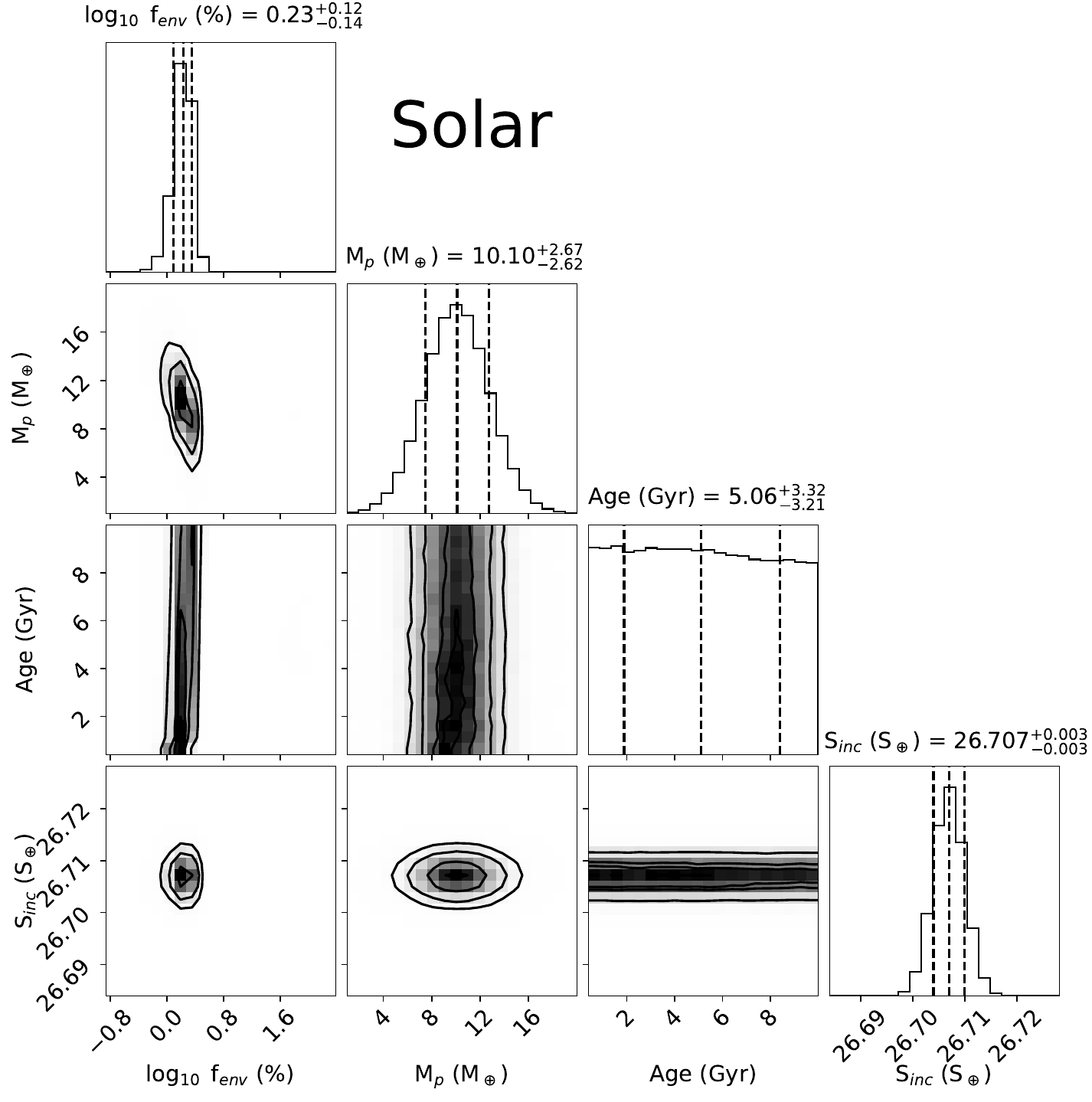} \includegraphics[width=0.24\textwidth]{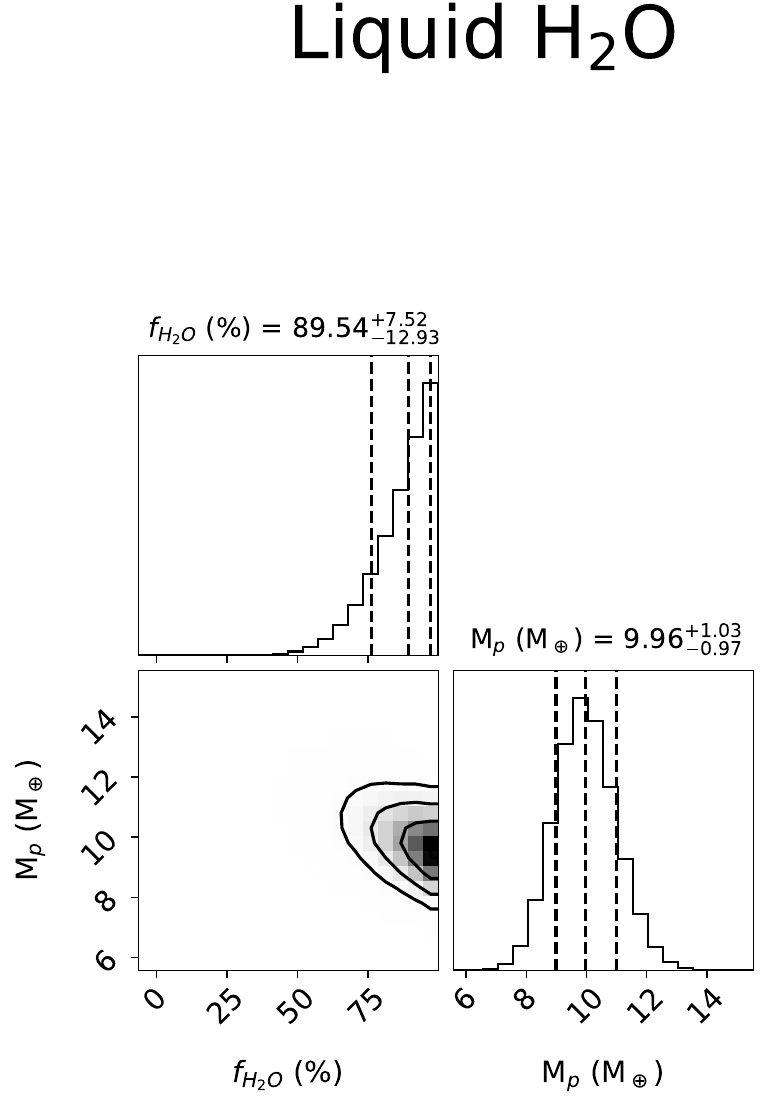}
\caption{Results from interior structure modeling with \texttt{smint}. Left panel: $f^{\prime}_{\mathrm{core}}$, $f_{\mathrm{H_2O}}$, $T_{\mathrm{eq}}$ and $M_{\mathrm{p}}$ for TOI-836c using the model from \cite{Aguichine2021}. Middle panel: $f_{\mathrm{env}}$, $M_{\mathrm{p}}$, age and $S_{\mathrm{inc}}$ for TOI-836c using the model from \cite{Lopez2014}. Right panel: $f_{\mathrm{H_2O}}$ and $M_{\mathrm{p}}$ for K2-18b using the model from \cite{Zeng2016}.}
\label{fig:corner-interior}   
\end{centering}
\end{figure*}

In this Appendix, we explain how we determine the bulk composition of planets, including TOI-836c, from interior structure models as presented in Section~\ref{sec:bulk}. For the bulk water content, we use the model from \citet{Aguichine2021}, where the planetary radius is a function of the planet mass, equilibrium temperature, reduced core mass fraction $f^{\prime}_{\mathrm{core}}$ and water mass fraction $f_{\mathrm{H_2O}}$. $f^{\prime}_{\mathrm{core}}$ designates the proportion of the refractory part of the planet made of iron alloy. Values of 0, 0.325 and 1 correspond to a pure mantle case, Earth-like case, and pure iron alloy, respectively. The \texttt{smint} code uses Gaussian priors on the mass and temperature, and produces posteriors of $f^{\prime}_{\mathrm{core}}$ and $f_{\mathrm{H_2O}}$ that match the planetary radius. The results for TOI-836c are shown in the left panel of Figure~\ref{fig:corner-interior}. For all other exoplanets, we perform a similar analysis but we fix $f^{\prime}_{\mathrm{core}}$ to 0.325, so that for planets that lie on the Earth-like composition curve we retrieve a value of $f_{\mathrm{H_2O}}$ close to 0 (the Earth value being $5\times 10^{-4}$). 

For the H$_2$-He content, we use the model from \cite{Lopez2014}, where the planetary radius is a function of the planet mass, incident flux $S_{\mathrm{inc}}$, age, and envelope mass fraction $f_{\mathrm{env}}$. The refractory part is always assumed to be of Earth-like composition. Gaussian priors are used for mass and incident flux and a flat prior is used for age to produce a posterior of $f_{\mathrm{env}}$ that match the planetary radius. The results for TOI-836c are shown in the middle panel of Figure~\ref{fig:corner-interior}. The same procedure is applied to all other exoplanets, and we retrieve $f_{\mathrm{env}}\simeq 0$ for planets that lie on the Earth-like composition curve.

The model of \cite{Aguichine2021} is adapted to sub-Neptunes that have equilibrium temperatures $T_{\mathrm{eq}}\ge 400$ K, so that water in the envelope is in supercritical state. K2-18b is an exceptionally cold exoplanet with $T_{\mathrm{eq}}=278$ K, so for this planet we use the interior model of \cite{Zeng2016} where water is in the condensed phase. In this model, the radius of the planet is a function of the planet mass and water mass fraction $f_{\mathrm{H_2O}}$. The core is assumed to be made of pure mantle. The results for K2-18b as a planet with liquid water are shown in the right panel of Figure~\ref{fig:corner-interior}. Liquid water clearly cannot account for the large radius of K2-18b, since its interior is compatible with a 100\% water interior, which is unphysical. If liquid water is present on K2-18b, it is most likely under a thick H$_2$-He atmosphere. K2-18b would then be a so-called hycean planet \citep{Madhu2023}. In the absence of a tabulated grid of interior models for hycean planets, we retain the computed value of 90\% of bulk water content for the interior of K2-18b.

\end{appendix}
\end{document}